\documentclass[a4paper,12pt]{article}
\usepackage{jheppub}
\usepackage{amsmath,amssymb}
\usepackage{xcolor}
\usepackage{color, colortbl}
\usepackage{cancel}
\usepackage{multirow} 
\usepackage{booktabs} 
\usepackage{makecell} 
\usepackage{mathrsfs}
\usepackage{graphicx}
\usepackage[compat=1.1.0]{tikz-feynman}
\usepackage{enumerate}
\usepackage{inputenc}
\usepackage[shortlabels]{enumitem}
\usepackage{xspace}
\usepackage{slashed}

\def\eDMEFT{e\textsc{DMeft}\xspace}
\def\dd{\mathrm{d}}
\def\micrOMEGAs{\texttt{micrOMEGAs}\xspace}
\def\MadGraph{\texttt{MadGraph5\_aMC@NLO}\xspace}
\def\MeV{\, \mathrm{MeV}}
\def\GeV{\, \mathrm{GeV}}
\def\TeV{\, \mathrm{TeV}}
\def\pb{\, \mathrm{pb}}
\def\fb{\, \mathrm{fb}}
\def\ab{\, \mathrm{ab}}

\author[a,b]{Giorgio Arcadi,}
\author[a,b,c,d]{David Cabo-Almeida,}
\author[e]{Sven Fabian,}
\author[e]{Florian Goertz}
\affiliation[a]{Dipartimento di Scienze Matematiche e Informatiche, Scienze Fisiche e Scienze della Terra,  Universita degli Studi di Messina,\\ Via Ferdinando Stagno d'Alcontres 31, I-98166 Messina, Italy}
\affiliation[b]{ INFN Sezione di Catania, Via Santa Sofia 64, I-95123 Catania, Italy}
\affiliation[c]{Departament de Física Quàntica i Astrofísica, Universitat de Barcelona,\\
Martí i Franquès 1, E08028 Barcelona, Spain}
\affiliation[d]{Institut de Ciències del Cosmos (ICCUB), Universitat de Barcelona,\\
Martí i Franquès 1, E08028 Barcelona, Spain}
\affiliation[e]{Max-Planck-Institut f{\"u}r Kernphysik\\ Saupfercheckweg 1, 69117 Heidelberg, Germany}
\emailAdd{giorgio.arcadi@unime.it}
\emailAdd{david.cabo@ct.infn.it}
\emailAdd{sven.fabian@mpi-hd.mpg.de}
\emailAdd{florian.goertz@mpi-hd.mpg.de}

\title{Dark Particles at the LHC:\\
{\large LHC-Friendly Dark Matter Characterization via Non-Linear EFT}}

\abstract{
In this work we illustrate a general framework to describe the LHC phenomenology of extended scalar (and fermion) sectors, with focus on dark matter (DM) physics, based on an effective field theory (EFT) with non-linearly realized electroweak symmetry. Generalizing Higgs EFT (HEFT), the setup allows to include a generic set of new scalar resonances, without the need to specify their UV origin, that could for example be at the interface of the Standard Model (SM) and the DM world.
In particular, we study the case of fermionic DM interacting with the SM via two mediators, each of which can possess either CP property and originate from various electroweak representations in the UV theory. Besides trilinear interactions between the mediators and DM or SM pairs (including pairs of gauge field-strength tensors), the EFT contains all further gauge-invariant operators up to mass dimension $D=5$. While remaining theoretically consistent, this setup offers enough flexibility to capture the phenomenology of many benchmark models used to interpret the results of experimental DM and BSM searches, such as two-Higgs doublet extensions of the SM or singlet extensions. Furthermore, the presence of two mediators with potentially sizable couplings allows to account for a broad variety of interesting collider signatures, as for example detectable mono-$h$ and mono-$Z$ signals. Correlations can be employed to diagnose the nature of the new particles.
}

\begin{document}
\maketitle

\section{Introduction}
With a clear hint on the nature of Dark Matter (DM) missing, it is of utmost importance to develop versatile theoretical frameworks for DM characterization in order not to miss potential signs of a dark sector. 
While various directions have been followed to this end, they all face their different challenges. For example, while an effective field theory (EFT) of DM, coupled to the Standard Model (SM) via higher dimensional operators, allows in principle for a very general parametrization of physics of dark sectors~\cite{Shepherd:2009sa,Beltran:2010ww,Goodman:2010ku,Bai:2010hh,Goodman:2010yf,DelNobile:2011uf,Bertuzzo:2024bwy}, such an approach is in general not applicable to LHC searches for DM. Since lacking a particle in the spectrum that mediates interactions between the SM and the DM that could go on-shell, one generically does not arrive at a detectable cross section for a consistent cutoff of the EFT, see e.g.~\cite{Busoni:2013lha,Bauer:2016pug}.

Thus, in order to use the synergy of different searches for DM, combining direct (and indirect) detection experiments with LHC searches, a different approach is required.
One possible avenue is to consider so-called simplified models (see, e.g.,~\cite{Abdallah:2015ter}), where a mediator is included. However, at least for scalar mediators, these models are in general no proper quantum field theories since they do not respect the SM gauge group in coupling the mediator to DM and to the SM, which puts also their renormalizability at stake.
Moreover, they could still miss important aspects of DM physics since they do not allow for additional new physics (NP) besides the single DM particle and one mediator. To resolve this dichotomy between generality/consistency and applicability at colliders, we seek a hybrid formulation that combines the virtues of both approaches, while avoiding their shortcomings. To this end, recently the `extended Dark Matter Effective Field Theory' (\eDMEFT) framework~\cite{Alanne:2017oqj,Alanne:2020xcb} was proposed. It keeps both the DM particle and the mediator between the dark sector and the SM as propagating degrees of freedom and thus remains valid at colliders, where DM EFT becomes problematic. 
On the other hand, retaining higher-dimensional operators ensures to have a (order-by-order) renormalizable field theory and to preserve gauge invariance, including induced correlations. Furthermore, the effect of additional NP~states, expected to be generally present in a realistic model of a dark sector, can be accounted for.

The framework presented in~\cite{Alanne:2020xcb} is still affected by some limitations. Indeed it strictly assumed the BSM mediator to the dark sector to be a SM singlet. Consequently, the interaction of the latter and the SM fermions were suppressed at least by one power of the effective NP  scale $\Lambda$, unless a mass mixing with the SM Higgs was assumed.  In a similar fashion, also the couplings with the gauge bosons could originate only from higher-dimensional~(hence suppressed) operators. Still, a viable DM phenomenology was possible in this framework and interesting collider phenomenology emerged in regions of the EFT parameter space~\cite{Alanne:2020xcb}, exceeding the reach of simplified models. It is also worth remarking that the so-called next-generation simplified models~\cite{Abe:2018bpo}, like, e.g., the 2HDM plus a scalar or pseudoscalar singlet with respect to the SM gauge group, are gathering increasing attention from the experimental community thanks to the broad variety of collider signatures they can account for. Along these lines, the purpose of this work is pushing the, more generic, \eDMEFT framework one step further. More precisely, two dynamical mediators will be considered in order to possibly incorporate the collider phenomenology of the aforementioned next-generation simplified models, with enhanced mono-X, signatures as well as UV-complete frameworks with various mediators, thereby capturing a broad class of dark sector models. The second relevant improvement is the implementation of the interactions of the mediators in an analogous way to the non-linear Higgs EFT~(HEFT), hence accounting for the possibility that the mediators belong to non-trivial $SU(2)$ representations. 
Similarly to~\cite{Alanne:2020xcb}, we focus on the case of spin-0 mediators with potentially different CP properties and fermionic DM. As will be discussed below, the framework is flexible enough to incorporate scalar DM as well. 

This article is organized as follows. In Section \ref{Sec:IntroEFT}, we set up the theory, being a non-linear sigma model in the phase of `broken' (i.e. non-linearly realized) electroweak~(EW) symmetry, and flash some relevant phenomenological applications that the framework can cover. We also discuss how a global analysis of the \eDMEFT can shed light on the UV nature of the dark sector. After that, in Section~\ref{Sec:IntroUV}, we match explicit SM extensions with additional scalars and fermions to the non-linear \eDMEFT and compare in detail the corresponding predictions for different observables to each other, including the relic abundance, direct-detection cross sections, and collider observables. Finally, in Section~\ref{sec:concl} we present our conclusions. 

\section{The extended DM effective field theory and applications} \label{Sec:IntroEFT}
We envisage an extended scalar sector, containing two new light degrees of freedom $({\cal S}_1,{\cal S}_2)^T$, being gauge singlets under $SU(3)_{c} \times U(1)_{\mathrm{em}}$, that could be embedded in various representations of $SU(2)_{L}$ in the UV theory. Moreover, they might either furnish the DM (via the lighter state, given it features a stabilizing symmetry), or connect the DM, which in this case we model as a $\mathbb{Z}_{2}$-odd fermionic singlet~$\chi$, to the SM.  To capture this general setup in a theoretically consistent way, we construct an extended EFT in the phase of non-linearly realized EW symmetry, following the HEFT~approach~\cite{deFlorian:2016spz}. The corresponding effective Lagrangian up to dimension~$D=5$ we consider in this work reads
\begin{align}
{\cal L} \supset &\frac 1 2 \sum_{\phi=\mathcal{S}_{1},\mathcal{S}_{2},h} \partial_{\mu} \phi \partial^{\mu} \phi- \mathcal{O}_{5}^{\lambda} + \frac{v^{2}}{4} \mathrm{Tr} \left[ \left( D_{\mu} \Sigma \right)^{\dagger}
\left( D^{\mu} \Sigma \right) \right] \mathcal{O}_{3}^{\kappa} \nonumber \\
&+ i\frac{v^{2}}{4} \mathrm{Tr} \left[\Sigma^{\dagger}
 \left( D^{\mu} \Sigma \right) \sigma^{3} \right] \left( \partial_{\mu} h\, \mathcal{O}_{2}^{h} + \partial_{\mu} \mathcal{S}_{1}\, \mathcal{O}_{2}^{s1} + \partial_{\mu} \mathcal{S}_{2}\, \mathcal{O}_{2}^{s2} \right) \nonumber \\
& - \frac{v}{\sqrt{2}} \left( \begin{pmatrix} \overline{u_{i,L}} && \overline{d_{i,L}} \end{pmatrix}
\Sigma \begin{pmatrix} Y_{ij}^{u} u_{j,R} \\ Y_{ij}^{d} d_{j,R} \end{pmatrix} \mathcal{O}_{2}^{c_{q}} + \begin{pmatrix} \overline{\nu_{i,L}} && \overline{\ell_{i,L}}  \end{pmatrix} \Sigma \tfrac{1-\sigma_3}{2}Y_{ij}^{\ell} \begin{pmatrix} {\nu_{j,R}} \\ {\ell_{j,R}}  \end{pmatrix} \mathcal{O}_{2}^{c_{\ell}} + \mathrm{h.c.} \right) \nonumber \\
&- \sum_{\phi}\frac{\phi}{16\pi^{2}}\left[g^{\prime 2} c_{B}^{\phi} B^{\mu\nu} B_{\mu\nu} + g^{2} c_{W}^{\phi} W^{I\mu\nu} W_{\mu\nu}^{I} 
+ g_{s}^{2} c_{G}^{\phi} G^{a\mu\nu} G_{\mu\nu}^{a} \right] \nonumber \\
&- \sum_{\phi}\frac{\phi}{16\pi^{2}}\left[ g^{\prime 2} \tilde{c}_{B}^{\phi} B^{\mu\nu} \widetilde{B}_{\mu\nu} + g^{2} \tilde{c}_{W}^{\phi} W^{I\mu\nu} \widetilde{W}_{\mu\nu}^{I} 
+ g_{s}^{2} \tilde{c}_{G}^{\phi} G^{a\mu\nu}\widetilde{G}_{\mu\nu}^{a}\right]\,,
\label{Eq:eDMEFT_Lagrangian}
\end{align}
with the chiral SM fermions~$f_{L,R}^{(j)} \equiv (1 \mp \gamma^{5})/2 f^{(j)}$ ($f=u,d,\nu,\ell$) and $Y^f_{ij}$ representing the Standard Model Yukawa couplings, with generation indices $i,j=1,2,3$. We assume the hypothesis of flavor alignment, leading to flavor-independent couplings $c_f$, even though a generalization to a general flavor structure is straightforward. The operators encoding all scalar combinations of the Higgs boson~$h$ and the additional scalars $\mathcal{S}_{i}$ up to dimension~$d$ are defined as
\begin{align}
\mathcal{O}_{d}^{C} \equiv {\cal O}_d^C(h,{\cal S}_1,{\cal S}_2) \equiv  \sum_{k=0}^d\sum_{j=0}^{d-k}\sum_{i=0}^j\, C^{(k)}_{i,j-i}\ h^k {\cal{S}}_1^i {\cal{S}}_2^{j-i} \, .
\end{align}	
Note that the effective operator $\mathcal{O}_{3}^{\kappa}$ is defined to contain at least one field, i.e.~$k+i+j>0$, as it contributes to gauge boson masses otherwise. 
The scalars $h,{\cal S}_i$ transform as $SU(2)_L$ singlets in the EFT, but, as discussed above, could be parts of larger multiplets in the UV theory, with the additional modes integrated out.
Gauge invariance in the couplings to non-trivial $SU(2)_L$ representations is achieved by dressing with the EW Goldstone matrix 
\begin{align}
\Sigma(x)= e^{i \sigma^{j} G^{j}(x)/v}\,,
\end{align}
with the SM-like Goldstones~$G^{j}$, that transforms as 
\begin{align}
\Sigma \to e^{i \varphi_{L}^{j}(x) \sigma^{j}/2}\, \Sigma\, e^{-i \varphi_{Y}(x) \sigma^{3}/2}
\end{align}
under the SM gauge symmetry and its covariant derivative reads 
\begin{align}
    D_\mu \Sigma \equiv \partial_{\mu}\Sigma -i\frac{g}{2} \sigma^{a}W_{\mu}^{a}\Sigma + i\frac{g^{\prime}}{2} B_{\mu} \Sigma\sigma^{3} \, .
\end{align}  
We thereby extend setups of non-linearly realized EW symmetry featuring a single {\it Higgs-like} particle in the EFT, i.e. HEFT (see \cite{Lee:1972yfa,Weinberg:1978kz,Feruglio:1992wf, Burgess:1999ha, Grinstein:2007iv, Contino:2010mh,Contino:2010rs, Alonso:2012px, Buchalla:2015wfa, Buchalla:2016bse, Falkowski:2019tft, Cohen:2020xca, Buchalla:2023hqk}), to include further scalar states from EW multiplets.
Their nature will have an impact on the power counting. For example, if $\mathcal{S}_{i}$ is part of an $SU(2)_L$ doublet, $c_{V}^{\mathcal{S}_{1,2}} $ will scale like $v/\Lambda^2$, with $v$ the EW vacuum expectation value (vev) and $\Lambda$ the EFT cutoff, while if it is a full singlet, the coupling will scale as $1/\Lambda$, and similarly for other coefficients, see the examples presented below. In particular, if the mediators are assumed to be SM singlets, as done in~\cite{Alanne:2020xcb}, this implies a suppression of the coefficients of their Yukawa-like couplings by a factor $1/\Lambda$. In the more flexible setup considered here this suppression factor will not necessarily be present since the scalar mediators can originate from, e.g., an $SU(2)$ doublet and consequently feature renormalizable couplings with the SM fermions. LHC signals would allow to discriminate the different completions via their scaling in the EFT.

In this work we will focus on fermionic DM that transforms as a singlet under the SM gauge group.\footnote{In the case of spin-0 mediators there is no substantial phenomenological difference between Dirac and Majorana DM.} The corresponding DM Lagrangian reads
\begin{align}
{\cal L}_\chi  = \overline{\chi} i \slashed{\partial} \chi 
  -{\cal O}_2^y(h,{\cal S}_1,{\cal S}_2)\, \overline{\chi_{L}} \chi_R +\mathrm{h.c.}\,.
\end{align}
An \emph{ad hoc} $\mathbb{Z}_{2}$ symmetry ensures the stability of the DM. Although we will focus on fermionic DM throughout this paper, we remark that scalar DM can be straightforwardly incorporated in our setup as well.\footnote{It could for example be interesting to study the matching of the scenario to inert-doublet-like models, that simultaneously allow to address the baryon asymmetry, see, e.g.~\cite{Astros:2023gda}.} Furthermore, the setup can also describe pseudoscalar mediators, such that terms with an odd number of pseudoscalars in 
$\mathcal{O}_{5}^{\lambda}(h,\mathcal{S}_{1},\mathcal{S}_{2})$ vanish, while terms with an even number of pseudoscalars in $\mathcal{O}_{2}^{h,s_1,s_2}$ vanish, in $\mathcal{O}_{2}^{c_f}(h,\mathcal{S}_{1},\mathcal{S}_{2})$ and $\mathcal{O}_{2}^{y}(h,\mathcal{S}_{1},\mathcal{S}_{2})$ corresponding coefficients become imaginary, and the field-strengths terms $\sim c_{V}^{\mathcal{S}_i}$ feature one dual $\widetilde V_{\mu \nu}$ instead of a tensor~$V_{\mu \nu}$. Lastly, the non-vanishing couplings in $\mathcal{O}_{3}^{\kappa}(h,\mathcal{S}_{1},\mathcal{S}_{2})$ that depend on the nature of the mediators~$\mathcal{S}_{1,2}$ can also be found. 

While the complexity of the presented framework is comparable to second-generation simplified models (such as the 2HDM+s/a setup), it is much more general. Due to its EFT character on the one hand and the less-constraining non-linearly realized symmetry (capturing various scalar embeddings) on the other, the \eDMEFT is capable of consistently describing the most relevant effects of a plethora of DM setups beyond the 2HDM+s/a and standard simplified models, including fermionic DM frameworks with scalar singlet mediator coupled to the SM for example via vector-like fermions~(see, e.g.,~\cite{Fan:2015sza,Gopalakrishna:2017zku,Alvarado:2017bax}), models with a composite scalar (Goldstone) mediator, the NMSSM with singlino DM and light scalar mediator, and multi-scalar mediator models with non-trivial cosmology, to just name a few.\footnote{We note that in some cases the non-linear DM EFT collapses to the (simpler) \eDMEFT~setup presented in~\cite{Alanne:2017oqj,Alanne:2020xcb,Goertz:2019vht}.} The non-linear \eDMEFT thereby provides a consistent UV completion and generalization of simplified models at the EW scale.

Contrary to what was done in~\cite{Alanne:2020xcb}, we will not comprehensively scan the parameter space of the non-linear \eDMEFT individually, but will rather be interested in a proof-of-principle of its capability of incorporating the phenomenology of a broad variety of complete models. To this end, we will compare different realizations of the \eDMEFT, featuring two scalar mediators, two pseudoscalar mediators, or one scalar and one pseudoscalar, to increasingly refined models.

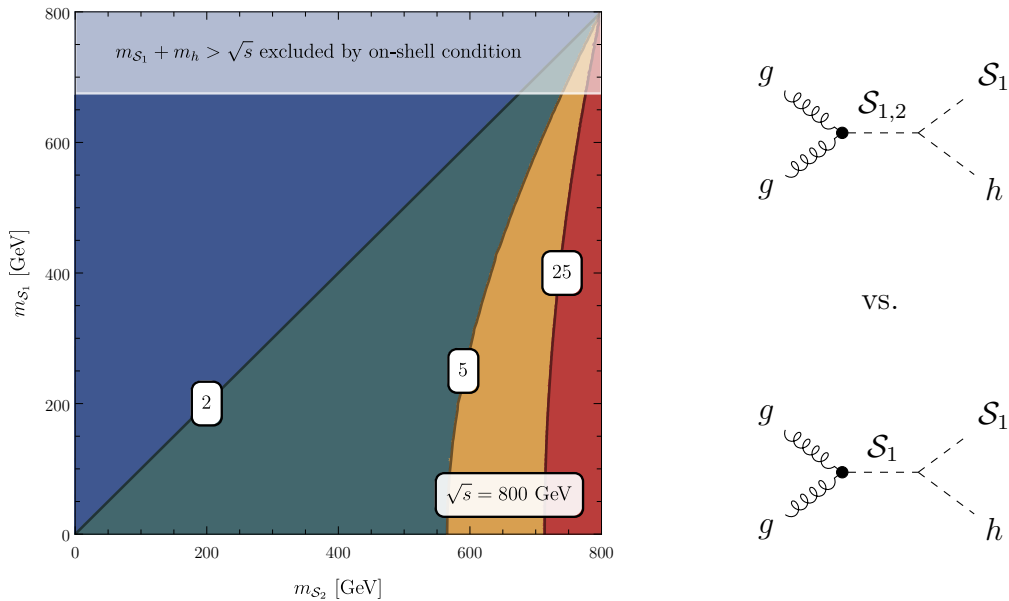
\begin{figure}[b!]
    \centering
    \begin{minipage}{0.59\textwidth}
    \centering
    \includegraphics[width=0.9\textwidth]{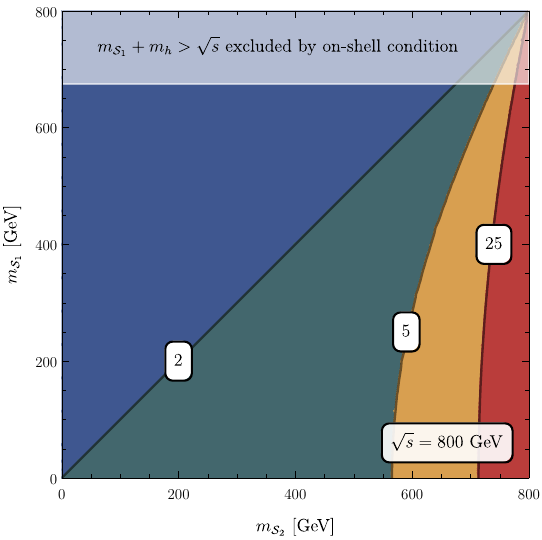}
    \end{minipage}
    \begin{minipage}{0.39\textwidth}
    \centering
    \begin{tikzpicture}[baseline=0mm]
        \begin{feynman}
            \vertex (a2) {\(g\)};
            \vertex[above=1.5cm of a2] (a1) {\(g\)};

            \vertex at ($(a1)!0.5!(a2) + (1cm,0)$)[dot] (b1) {};
            \vertex[right=1cm of b1] (c1);

            \vertex at ($(a1) + (3cm,0)$) (d1) {\(\mathcal{S}_{1}\)};
            \vertex at ($(a2) + (3cm,0)$) (d2) {\(h\)};

            \vertex[below=3cm of a2] (a3) {\(g\)};
            \vertex[below=1.5cm of a3] (a4) {\(g\)};

            \vertex at ($(a3)!0.5!(a4) + (1cm,0)$)[dot] (b2) {};
            \vertex[right=1cm of b2] (c2);

            \vertex at ($(a3) + (3cm,0)$) (d3) {\(\mathcal{S}_{1}\)};
            \vertex at ($(a4) + (3cm,0)$) (d4) {\(h\)};
        
            \diagram* {
            (a1) -- [gluon] (b1),
            (a2) -- [gluon] (b1),
            (b1) -- [scalar,edge label=\(\mathcal{S}_{1,2}\)] (c1),
            (c1) -- [scalar] (d1),
            (c1) -- [scalar] (d2),

            (a3) -- [gluon] (b2),
            (a4) -- [gluon] (b2),
            (b2) -- [scalar,edge label=\(\mathcal{S}_{1}\)] (c2),
            (c2) -- [scalar] (d3),
            (c2) -- [scalar] (d4),
            };
        \end{feynman}
        \node[] at (1.5cm,-1.5cm) {$\mathrm{vs.}$};    
    \end{tikzpicture}
    \end{minipage}
    \caption{Comparison of the cross sections for the gluon fusion process $gg \rightarrow \mathcal{S}_{1}h$, contributing to the mono-$h$ signature, in an \eDMEFT with two scalar mediators~(present work) and one mediator~(cf. Ref.~\cite{Alanne:2017oqj}). \emph{Left}: Ratio of the cross sections,~$\sigma_{2\mathrm{med.}}/\sigma_{1\mathrm{med.}}$, in terms of the mediator masses~$m_{\mathcal{S}_{1,2}}$ and for the trilinear couplings assumed to be equal. \emph{Right}: Diagrams taken into account, where the upper ones correspond to the new non-linear \eDMEFT, while only the lower contributes in the single-mediator setup (similar to standard simplified models).}
    \label{Fig:eDMEFT_Crossx}
\end{figure}

Before doing so, we however close this section by exploring quantitatively the possible enhancement in mono-X signatures compared to simplified models and the original \eDMEFT. Therefore, in Fig.~\ref{Fig:eDMEFT_Crossx}, we compare the cross section for the production of the DM mediator $\mathcal{S}_1$ in association with the Higgs boson, contributing to the mono-$h$ process, in the non-linear two-mediator \eDMEFT ($\sigma_{2\mathrm{med.}}$) to the one-mediator case ($\sigma_{1\mathrm{med.}}$). For simplicity, we turn on just the couplings of the mediators to gluons and the trilinear interactions between two mediators and the Higgs boson, assuming equal coupling strength, which is enough to make the point. Moreover, we work with a fixed partonic center-of-mass energy of $\sqrt s =800\GeV$, even though in a realistic hadron-collider environment this would need to be integrated over, weighted with parton distribution functions, which would just make the effects stronger.

Inspecting the left panel of Fig.~\ref{Fig:eDMEFT_Crossx}, we can observe a sizable enhancement of the cross section in the two-mediator \eDMEFT, reaching up to (several) orders of magnitude -- as indicated by the contour lines of constant $\sigma_{2\mathrm{med.}}/\sigma_{1\mathrm{med.}}$. This is driven by the possibility of the second meditator $\mathcal{S}_2$ to go (approximately) on shell in the s-channel, while simultaneously the emitted $\mathcal{S}_{1}$ and $h$ can be on shell given that $m_{\mathcal{S}_{2}}\geq m_{\mathcal{S}_{1}}+m_h$. In the one-mediator case, such a resonant enhancement is not possible. The corresponding diagrams are displayed in the upper-right and lower-right panel of~Fig.\ref{Fig:eDMEFT_Crossx}, respectively.
\section{Matching with UV models}\label{Sec:IntroUV}

In this section we will compare the non-linear \eDMEFT to a selection of complete models in order to envisage its capability of being a flexible interface between theory and experiments. This comparison will always proceed along the same steps. For each complete model we will identify the states corresponding to $\mathcal{S}_{1,2}$ and the heavy degrees of freedom residing at the NP scale~$\Lambda$ and thus being integrated out from the particle spectrum in the non-linear \eDMEFT. Parameter scans in the UV model will be conducted, the matching conditions as the relations between the Wilson coefficients and the UV model parameters determined, and ultimately used to compute the corresponding parameter points in the considered \eDMEFT realization~(cf.~Appendix~\ref{App:TableWilsonCoefficients} for relevant Wilson coefficients). The following three classes of observables will be discussed: the DM relic abundance, computed by using the public package \micrOMEGAs~\cite{Belanger:2013oya,Belanger:2018ccd} and assuming the standard freeze-out paradigm; the DM scattering cross section in direct-detection experiments; and ultimately a selection of collider signals with particular focus on the mono-$h$ and mono-$Z$ signatures. For the latter, we compute the production cross section of these signatures with the \MadGraph code~\cite{Alwall:2011uj,Alwall:2014hca}.
We consider either the relative difference or the ratio of the predictions of the concrete model and the \eDMEFT for each of the aforementioned observables. The lower the relative difference or the closer the ratio to unity, the more accurately the \eDMEFT captures the phenomenology of the considered concrete model.

\subsection{SM + one complex scalar singlet}

The first~(and simplest) model to be considered here features a scalar sector composed of the SM Higgs doublet~$\Phi$ and a complex scalar~$S$ which transforms trivially under the SM gauge group but non-trivially under a global, spontaneously broken~$U(1)$. The two fields can be decomposed as
\begin{align}
    \Phi = \begin{pmatrix}
        G^{+} \\ \left( v_{h} + \hat{h} + i G^{0} \right)/\sqrt{2} 
    \end{pmatrix} \quad , \quad S = \frac{1}{\sqrt{2}} (v_{s} + \hat{s} + i a)\,,
\end{align}
with the would-be Goldstones~$G^{\pm,0}$. The hat notation indicates that these scalars are not mass eigenstates. The scalar potential reads
\begin{align}
    -\mathcal{L} &\supset \mu_{\Phi}^{2} \left\vert \Phi \right\vert^{2} + \lambda_{\Phi} \left\vert \Phi \right\vert^{4} + \mu_{S}^{2} \left\vert S \right\vert^{2} + \lambda_{S} \left\vert S \right\vert^{4} + \lambda_{\Phi S} \left\vert \Phi \right\vert^{2} \left\vert S \right\vert^{2} + \mu_{a}^{2} a^{2}\,,
\end{align}
with the last term explicitly breaking the global~$U(1)$ and hence generating the parameterically small mass of the pseudo-Goldstone boson~$a$. The loop-induced decay of the pseudo-Goldstone into gauge bosons (e.g. photons) prevents it from being cosmologically stable. 

In presence of a non-zero~$\lambda_{\Phi S}$, there is a mass mixing between the CP-even components of $\Phi$ and $S$, described by the following mass matrix
\begin{align}
    \mathcal{M}_{\mathrm{s}} = \begin{pmatrix}
2 v_{h}^{2} \lambda_{\Phi} & \lambda_{\Phi S} v_{h} v_{s} \\
\lambda_{\Phi S} v_{h} v_{s} & 2 v_{s}^{2} \lambda_{S}
    \end{pmatrix} \, .
\end{align}
Upon diagonalization, the masses of the SM-like Higgs and the BSM scalar are given by
\begin{align}
    m_{h,s}^{2} &= v_{h}^{2} \lambda_{\Phi} + v_{s}^{2} \lambda_{S} \pm  \frac{v_{h}^{2} \lambda_{\Phi} - v_{s}^{2} \lambda_{S}}{\cos 2 \theta}\,,
\end{align}
with $\theta$ the mixing angle between the scalars, while the mass of the pseudoscalar is $m_{a} \simeq \mu_{a}$.

Furthermore, we consider four heavy chiral quarks~$\mathcal{B}_{L,R}$, $\mathcal{T}_{L,R}$ and two chiral DM fermions~$\chi_{L,R}$. The heavy chiral quarks are charged under the $SU(3)_{c} \times U(1)_{\rm em}$ like their SM siblings, while the DM fermion is a SM singlet. The BSM fermions transform non-trivially under the global~$U(1)$ such as to allow for interactions with the BSM scalar singlet~$S$.
The representations for the BSM fields under $\mathrm{SM}\times U(1)$ are presented in Tab.~\ref{Tab:MonoSingletModel_QuantumNumbers} and give rise to
\begin{table}[b!]
    \centering
    \begin{tabular}{ccccc}
        \toprule
        \multicolumn{1}{c}{Field} & \multicolumn{1}{c}{$SU(3)_{c}$} & \multicolumn{1}{c}{$SU(2)_{L}$} & \multicolumn{1}{c}{$U(1)_{Y}$} & \multicolumn{1}{c}{global $U(1)$} \\
        \midrule
        $S$ & $\mathbf{1}$ & $\mathbf{1}$ & $0$ & $+1$ \\
        $\mathcal{T}_{L}$ & $\mathbf{3}$ & $\mathbf{1}$ & $4/3$ & $+1/2$ \\
        $\mathcal{T}_{R}$ & $\mathbf{3}$ & $\mathbf{1}$ & $4/3$ & $-1/2$ \\
        $\mathcal{B}_{L}$ & $\mathbf{3}$ & $\mathbf{1}$ & $-2/3$ & $+1/2$ \\
        $\mathcal{B}_{R}$ & $\mathbf{3}$ & $\mathbf{1}$ & $-2/3$ & $-1/2$ \\
        $\chi_{L}$ & $\mathbf{1}$ & $\mathbf{1}$ & $0$ & $+1/2$ \\
        $\chi_{R}$ & $\mathbf{1}$ & $\mathbf{1}$ & $0$ & $-1/2$ \\
        \bottomrule
    \end{tabular}
    \caption{Quantum numbers of BSM fields in complex scalar-singlet extension.}
    \label{Tab:MonoSingletModel_QuantumNumbers}
\end{table}
\begin{align}
    -\mathcal{L} \supset y_{\chi S} \overline{\chi_{L}} S \chi_{R} + \sum_{\mathcal{Q}} y_{\mathcal{Q} S} \overline{\mathcal{Q}_{L}} S \mathcal{Q}_{R} + \mathrm{h.c.}\,,
\end{align}
with the heavy chiral quarks~$\mathcal{Q}=\mathcal{B},\mathcal{T}$ and the real coupling parameters~$y_{\chi S}$, $y_{\mathcal{Q} S}$. Note that the $U(1)$~charges prevent the BSM quarks from mixing with the SM quarks. 

The mass of the lightest BSM quark, which is generated after spontaneous symmetry breaking of the global~$U(1)$, corresponds to the cutoff scale~$\Lambda$ of the \eDMEFT. 
The DM mass can be much smaller than the cutoff scale,~i.e.~$m_{\chi} = y_{\chi S} v_{s}/\sqrt{2} \ll \Lambda$, for~$y_{\chi S} \ll 1$, while~$y_{\mathcal{Q} S} \gtrsim 1$ to ensure the high mass of the BSM quarks~$\mathcal{Q}$.
Integrating out the chiral fermions generates the effective couplings in Eq.~(\ref{Eq:eDMEFT_Lagrangian}) of the scalar and pseudoscalar components of $S$ with the gluons and the photons~\cite{Choi:2021nql}. We just present, for simplicity, the explicit expression for the effective coupling with the gluons (the other one can be straightforwardly obtained), reading
\begin{align}
    c_{G}^{s(h)} &= -\sum_{\mathcal{Q}} \frac{\,y_{\mathcal{Q}S} c_{\theta} (s_{\theta})}{2\,\sqrt{2}\, m_{\mathcal{Q}}} F_{S}\left( \frac{m_{s(h)}^{2}}{4 m_{\mathcal{Q}}^{2}} \right) \\ 
    c_{G}^{a} &= -\sum_{\mathcal{Q}} \frac{\,y_{\mathcal{Q}S}}{2\,\sqrt{2}\,m_{\mathcal{Q}}} F_{PS}\left( \frac{m_{a}^{2}}{4 m_{\mathcal{Q}}^{2}}\right) \, ,
    \label{Eq:MonoSingletModel_effectiveGluonCouplings}
\end{align}
where we have used the short-hand notations $s_{\theta} \equiv \sin\theta$ and $c_{\theta} \equiv \cos \theta$, and the $\mathcal{Q}$-loop functions are given by
\begin{align}
    F_{S} \left( x \right) &= \frac{1}{ x^{2}} \left( x + \left( x - 1 \right) \arcsin^{2} \sqrt{x} \right) \quad , \quad F_{PS} \left( x \right) = \frac{1}{x} \arcsin^{2} \sqrt{x} \, .
    \label{Eq:MonoSingletModel_LoopFunctions}
\end{align}
Due to the $s/h$ mixing, the effective coupling of the Higgs boson with gluons and photons get modified. Consequently, one has to ensure that the mixing angle is sufficiently small to avoid conflict with the experimental fit of the Higgs-signal strengths. The further coefficients in Eq.~(\ref{Eq:eDMEFT_Lagrangian}) can be obtained straightforwardly from the Lagrangians above, identifying $s,a$ with $\mathcal{S}_{1,2}$.

In the following subsections, this model will be confronted with the \eDMEFT featuring a scalar and a pseudoscalar mediator. To this end, a parameter scan has been conducted over the following ranges:
\begin{align}
    &m_{s} \in \left[ m_{h}/2, \Lambda \right] \quad , \quad m_{a} \in \left[ 1\GeV, 0.01m_{h} \right] \quad , \quad m_{\chi} \in \left[ m_{h}/2, \Lambda \right] \quad , \nonumber \\
    &v_{s} \in \left[ 100\GeV , 1600\GeV  \right] \quad , \quad \sin\theta \in \left[ 0,1 \right] \, .
    \label{Eq:MonoSingletModel_ParameterRanges}
\end{align}
As already pointed out, the scale $\Lambda$ corresponds to the mass of the BSM fermions, different from the DM candidate. 
Unless stated differently, we will set $\Lambda=1\TeV$. The most important theoretical and experimental constraints on the model are summarized in Appendix~\ref{App:OneSingletExtension_Constraints}. We already point out nevertheless the existence of strong constraints from invisible decays of the Higgs boson, arising when $h\rightarrow\chi \chi$, $h\rightarrow ss$ and/or $h\rightarrow aa$ become kinematically allowed. For this reason we have taken for simplicity $m_{h}/2$ as the boundary for the scan of both $m_{\chi}$ and $m_{s}$. The pseudoscalar state $a$ is instead always lighter than $m_h/2$. Consequently, we have imposed in the parameter scan $\mathrm{BR}(h \rightarrow aa)\leq 0.11$ and retained only the parameter assignations complying with this condition.
As a final remark, we point out the choice of very low values of $m_{a}$ is justified by the fact that $a$ is a pseudo-Goldstone boson with a very small mass originating by a tiny explicit breaking of the $U(1)$~symmetry.

\subsubsection{Dark Matter relic abundance}

First, we check whether the \eDMEFT can reproduce the phenomenology related to the DM relic density. Assuming the freeze-out paradigm, the latter is due to DM annihilation processes into SM pairs, via $s$-channel exchange of the $h/s/a$ scalars, as well as into scalar/pseudoscalar pairs via $t$-channel exchange of the DM. Furthermore, annihilations into $gg/\gamma \gamma/Z\gamma/ZZ$ can proceed via the effective $s/a$ couplings induced by the BSM fermions. The latter are not directly involved in the generation of the DM relic density as long as they are heavier than the DM. We have computed the DM relic density for the model points of the parameter scan described in the previous subsection and compared the obtained values to the ones derived in the matched \eDMEFT, taking the same values for the DM and mediator masses and couplings.

\begin{figure}[b!]
    \centering    \includegraphics{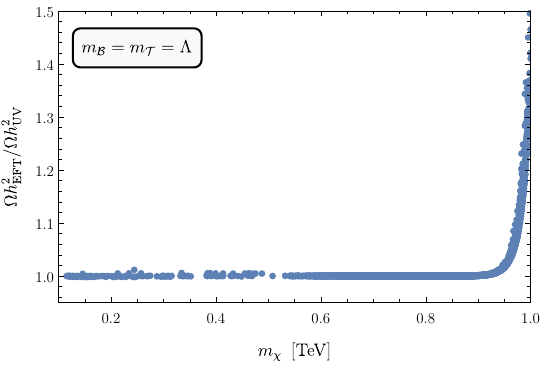}
    \caption{Ratio of the DM relic abundance in the \eDMEFT and the one in the mono-singlet model. Close to the new-physics scale~$\Lambda$ the heavy degrees of freedom impact the DM annihilation cross section and obviously the \eDMEFT cannot accurately describe the model any longer.}
    \label{Fig:MonoSingletModel_RatioRelicAbundance}
\end{figure}

Fig.~\ref{Fig:MonoSingletModel_RatioRelicAbundance} displays the ratio of the relic density derived in the EFT and in the complete model, as a function of the DM mass. 
The \eDMEFT reproduces very well the phenomenology of the mono-singlet model, given that the two models feature the same amount of $s$-channel mediators for DM interactions such that a successful matching is possible. Differences among the two models appear only as the DM mass approaches the scale $\Lambda$. In such a regime, the heavy degrees of freedom (here the heavy chiral quarks~$\mathcal{B},\mathcal{T}$) start to contribute to the DM annihilation processes as additional particles in the final state. Obviously, the DM relic abundance in the mono-singlet model decreases when the center-of-mass energy of the initial state opens further parameter space for DM annihilation. As expected, the EFT ceases to be valid in this case of the exchanged energy approaching the cutoff.

\subsubsection{Dark Matter Direct Detection}
The Dark Matter Direct Detection (DMDD) processes for this model are dominated by a tree-level $s/h$ exchange due to their mixing with each other.\footnote{The pseudoscalar enters in loop-induced interactions. The corresponding contributions are negligible with respect to the ones reported in the main text unless cancellations occur, like, for example, when $m_h \sim m_s$. Interactions between the DM and pseudoscalar mediator will, on the contrary, become important for the models considered in one of the upcoming sections and, consequently will be discussed in detail in that context.} The corresponding effective Lagrangian is given by
\begin{align}
\mathcal{L}_{\mathrm{eff}}= & \frac{1}{2} \sum_{q=u, d, s} C_q m_q \overline{\chi} \chi \overline{q} q - \frac{9 \alpha_s}{16 \pi} C_{G} \overline{\chi} \chi G_{\mu \nu}^{a} G^{a \, \mu \nu}
\label{Eq:DMDDeffectiveLagrangian}
\end{align}
with the Wilson coefficients
\begin{align}
    C_{q} &= \frac{ y_{\chi S} \sin 2\theta \left( m_{h}^{2} - m_{s}^{2} \right)}{\sqrt{2} m_{h}^{2} m_{s}^{2} v_{h}} \nonumber\\
    C_{G} &= \sum_{Q=c,b,t}\frac{2}{27}C_{Q} + \frac{ y_{\chi S}  \left( \sin\theta c_{G}^{h}\,m_{s}^{2} + \cos\theta c_{G}^{s}\,m_{h}^{2} \right)}{9\,\sqrt{2}\,m_{h}^{2} m_{s}^{2}}
\end{align}
and the Wilson coefficients~$c_{G}^{h,s}$ from Eq.~(\ref{Eq:eDMEFT_Lagrangian}). The cross section for DM scattering off nucleons hence reads
\begin{align}
    \sigma_{\chi N}^{\rm SI}=\frac{\mu_{\chi N}^2}{\pi}\left \vert C_N \right \vert^2
    \label{Eq:DMDDSIcrossSection}
\end{align}
with the DM-nucleon reduced mass~$\mu_{\chi N} = m_{N} m_{\chi} / (m_{N} + m_{\chi})$ and 
\begin{align}
C_{N} = m_{N}\left[ \sum_{q=u, d, s} C_{q} f_{T_{q}}^{N} + C_{G} f_{T_{G}}^{N} \right] \, .
\end{align}
The reader is referred to~\cite{Belanger:2013oya,Abe:2018emu} for the form factors $f_{T_{q}}^{N},f_{T_{g}}$ in the latter formula. Given the fact that the mediators of the \eDMEFT and the complete model match and being the typical energy scale of DMDD interactions very low, of the order of $1\GeV \ll \Lambda$, the \eDMEFT is always capable of capturing the DMDD phenomenology of the mono-singlet model.

\subsubsection{Collider signatures}
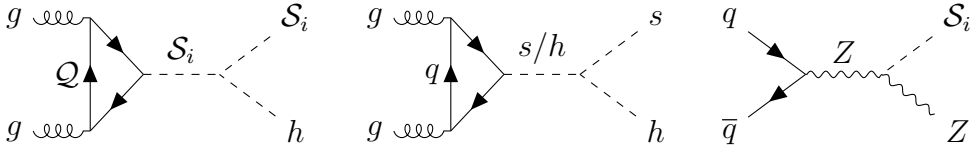
\begin{figure}[b!]
    \begin{tikzpicture}[baseline=7mm]
        \begin{feynman}
            \vertex (a1) {\(g\)};
            \vertex[above=1.5cm of a1] (a2) {\(g\)};

            \vertex[right=1cm of a1] (b1);
            \vertex[right=1cm of a2] (b2);

            \vertex at ($(b1)!0.5!(b2) + (0.7cm,0)$) (c);
            \vertex[right=1cm of c] (d);

            \vertex at ($(a1) + (3.7cm,0)$) (e1) {\(h\)};
            \vertex at ($(a2) + (3.7cm,0)$) (e2) {\(\mathcal{S}_{i}\)};

            \diagram* {
                (a1) -- [gluon] (b1),
                (a2) -- [gluon] (b2),
                (b1) -- [fermion,edge label=\(\mathcal{Q}\)] (b2),
                (b2) -- [fermion] (c),
                (c) -- [fermion] (b1),
                (c) -- [scalar, edge label=\(\mathcal{S}_{i}\)] (d),
                (e1) -- [scalar] (d),
                (e2) -- [scalar] (d)
            };
        \end{feynman}
    \end{tikzpicture}
    \hspace{2mm}
    \centering
    \begin{tikzpicture}[baseline=7mm]
        \begin{feynman}
            \vertex (a1) {\(g\)};
            \vertex[above=1.5cm of a1] (a2) {\(g\)};

            \vertex[right=1cm of a1] (b1);
            \vertex[right=1cm of a2] (b2);

            \vertex at ($(b1)!0.5!(b2) + (0.7cm,0)$) (c);
            \vertex[right=1cm of c] (d);

            \vertex at ($(a1) + (3.7cm,0)$) (e1) {\(h\)};
            \vertex at ($(a2) + (3.7cm,0)$) (e2) {\(s\)};

            \diagram* {
                (a1) -- [gluon] (b1),
                (a2) -- [gluon] (b2),
                (b1) -- [fermion,edge label=\(q\)] (b2),
                (b2) -- [fermion] (c),
                (c) -- [fermion] (b1),
                (c) -- [scalar, edge label=\(s/h\)] (d),
                (e1) -- [scalar] (d),
                (e2) -- [scalar] (d)
            };
        \end{feynman}

    \end{tikzpicture}
    \hspace{2mm}
    \centering
    \begin{tikzpicture}[baseline=7mm]
        \begin{feynman}
            \vertex (j1) {\(\overline{q}\)};
            \vertex[above=1.5cm of j1] (j2) {\(q\)};

            \vertex at ($(j1)!0.5!(j2) + (1cm,0)$) (k);
            \vertex[right=1cm of k] (l);

            \vertex at ($(j1) + (3cm,0)$) (m1) {\(Z\)};
            \vertex at ($(j2) + (3cm,0)$) (m2) {\(\mathcal{S}_{i}\)};
        
            \diagram* {
            (k) -- [fermion] (j1),
            (j2) -- [fermion] (k),
            (k) -- [boson,edge label=\(Z\)] (l),
            (l) -- [boson] (m1),
            (m2) -- [scalar] (l),
            };
        \end{feynman}
    \end{tikzpicture}
    \caption{Processes with gluons in the initial state for mono-$h$~(first two diagrams) and mono-$Z$ signatures with and without scalar resonance at hadron colliders in the complex-singlet model. The mediators are~$\mathcal{S}_{i} = s,a$. Note that the coupling of the pseudoscalar~$a$ to $Z$ in the third diagram involves a loop. 
    } 
    \label{Fig:MonoSingletModel_ColliderSignatures}
\end{figure}

\begin{figure}[b!]
    \centering
    \includegraphics[width=0.49\textwidth]{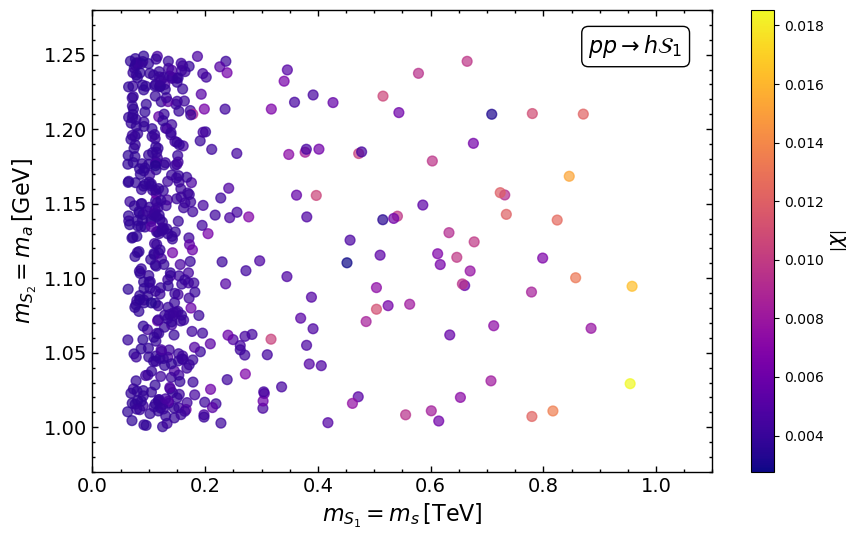}
    \includegraphics[width=0.49\textwidth]{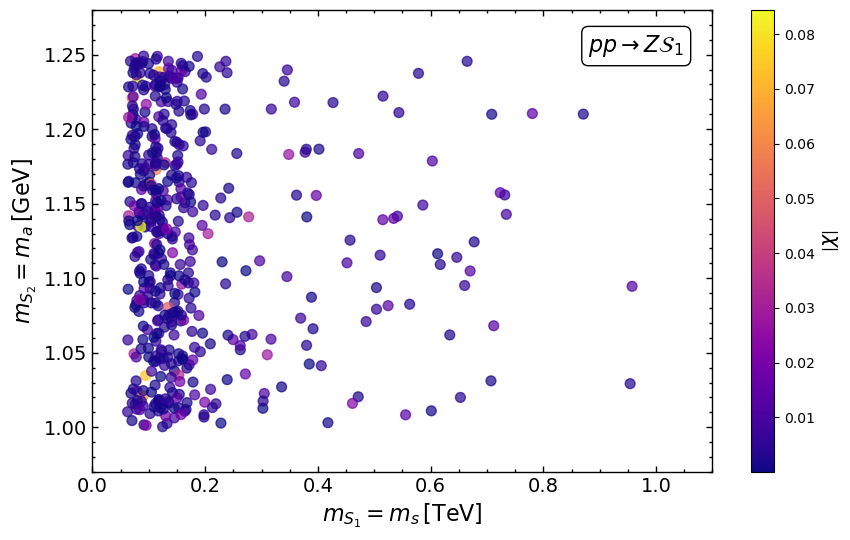}
    \includegraphics[width=0.49\linewidth]{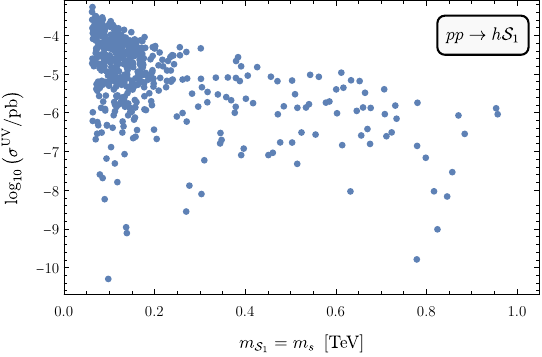}
    \includegraphics[width=0.49\linewidth]{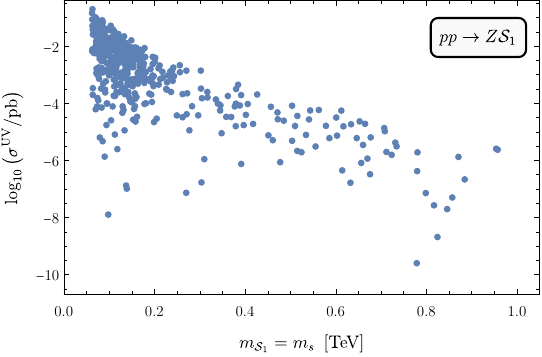}
    \caption{Comparison of the mono-$h$ (\emph{upper left}) and the mono-$Z$ (\emph{upper right}) production cross sections in the complex-singlet model and the \eDMEFT, where $\chi$ denotes the relative difference, and the absolute values of the cross sections~(\emph{lower row}). 
    }
    \label{Fig:MonoSingletModel_Crossx}
\end{figure}

As the last aspect of our comparison of the complex-singlet model and our \eDMEFT, we shall turn to appealing collider signatures at the LHC. For this sake we restrict ourselves to mono-$h$ and mono-$Z$ processes in $pp$~collisions, which are a subset of the potentially promising processes compiled in Appendix~\ref{App:TableWilsonCoefficients} and depicted in Fig.~\ref{Fig:MonoSingletModel_ColliderSignatures} for the UV model. Note that the annihilation of the SM quarks from the protons can be mediated by the photon as well, but this requires a BSM quark loop in the coupling to the scalars.
The mono-$h$~(left) and mono-$Z$~(right) production cross sections obtained in a parameter scan are shown in Fig.~\ref{Fig:MonoSingletModel_Crossx}.
For each parameter point the cross section has been computed as well for the \eDMEFT counterpart (see  below). Within the considered scan ranges, the maximal values of the mono-$h$ and mono-$Z$ cross sections for the UV models have been found to be (see the lower panel in Fig~\ref{Fig:MonoSingletModel_Crossx}):
\begin{align}
    \sigma_{\mathrm{max}} = \begin{cases}
        5.3 \times 10^{-4}\pb & \mathrm{for \ mono\text{-}}h \\
        2.0 \times 10^{-1} \pb & \mathrm{for \ mono\text{-}}Z
    \end{cases} \, .
\end{align}

The reason of the large suppression of the mono-$h$ process can be understood by looking at the Feynmann diagrams shown in Fig.~\ref{Fig:MonoSingletModel_ColliderSignatures}. Assuming that CP is preserved in the scalar sector, the mono-$h$ signature requires the trilinear coupling with two CP-even or two CP-odd states. Since the model under scrutiny features, in addition to the Higgs boson, a single CP even and a single CP odd state, one immediately concludes that the mono-$h$ signature cannot be associated to a resonant process and, consequently, the expected cross section is small. In addition one has to remember that, given the small Yukawa couplings of light fermions, the scalars are predominantly produced via gluon fusion, which is a loop induced process. While also the mono-$Z$ signature cannot be associated to resonance processes, the production cross section is considerably bigger as the $Z$-boson can be produced via quark fusion.\footnote{A resonant process might actually occur via a $saZ$ coupling. However the latter is zero (at tree level) in the mono singlet model as the new states are $SU(2)$ singlets.}

Assuming the final state bosons to be on-shell, the number of events associated to the signatures under consideration can be estimated via:
\begin{align}
    N = \sigma \times \mathcal{L}_{\mathrm{int}}  \times \varepsilon \times \mathrm{BR}\left( s \rightarrow \chi \overline{\chi} \right) \times \begin{cases}
        \mathrm{BR}\left( h \rightarrow b \overline{b} \right) &\mathrm{for \, mono\text{-}}h \\
        \mathrm{BR}\left( Z \rightarrow \ell \overline{\ell} \right) &\mathrm{for \, mono\text{-}}Z
    \end{cases}
    \label{Eq:NumberOfEventsAtColliders}
\end{align}
with the integrated luminosity~$\mathcal{L}_{\mathrm{int}} = 139 \fb^{-1}$~$(3\ab^{-1})$ for (HL-)LHC and the detector and analysis efficiency~$\varepsilon$. We have not explicitly considered the BSM pseudoscalar~$a$ as a final state, as it is very light and predominantly decays into DM pairs off-shell, further suppressing the cross section.
The SM branching ratio reads $\mathrm{BR}(h \rightarrow b \overline{b}) = 5.82\times 10^{-1}$~\cite{ParticleDataGroup:2022pth}, while the branching ratio of the BSM process~$s \rightarrow \chi \overline{\chi}$ depends on the masses~$m_{s,\chi}$, the mixing angle~$\theta$, and the Yukawa coupling~$y_{\chi S}$. Hence, the branching ratio is not constant throughout our set of parameter points. 

Exploiting the fact that neither the branching ratio nor the efficiency~$\varepsilon$ can be larger than unity, the number of mono-$h$~events in current LHC data is estimated not to exceed~$\sim \mathcal{O}(10)$. The number of events after the runtime of HL-LHC with the integrated luminosity stated above is not larger than~$\sim 10^{3}$. For the mono-$Z$ events, on the other hand, one can expect~$\mathcal{O}(10^{4(5)}) \times \mathrm{BR}(Z \rightarrow \ell \overline{\ell})$ events for the \mbox{(HL-)LHC} with the optimistic assumption of the largest possible detector and analysis efficiencies~$\varepsilon=1$.

We close with a comparison of the UV model and the \eDMEFT. The relative difference between their prediction, defined as $\chi$, characterizes the color code of Fig.~\ref{Fig:MonoSingletModel_Crossx}. As evident, the parameter $\chi$ is always in the few per cent range, demonstrating the capability of the \eDMEFT of reproducing the concrete model. The BSM chiral fermions can successfully be integrated out and enter only in the effective couplings of the mediators with the gluons, while the \eDMEFT scalars $\mathcal{S}_{1,2}$ can consistently be matched to the components of the complex singlet. As expected, the lighter the mediator the better the effective interactions describe the UV model.

\subsection{SM + two scalar singlets} \label{Sec:MultiSingletModel}
The model considered in the previous section allowed for a rather straightforward comparison to the \eDMEFT as the number of extra scalar degrees of freedom coincided with the ones present in the \eDMEFT. The only states integrated out have been the BSM chiral fermions, leading to the effective coupling of the BSM scalars with the gauge bosons. This feature is easily captured by the \eDMEFT due to the high masses of the chiral fermions. In this subsection we study a slightly more complicated model with a richer scalar sector. The comparison between the EFT and the complete model hence relies on the possibility of integrating out also degrees of freedom in the mediator sector in the latter. 
 We consider a scalar sector consisting of a complex scalar~$S_{1}$ and one real scalar~$S_{2}$ in addition to the SM doublet~$\Phi$, both transforming as singlets under the SM gauge group. The Lagrangian features a global $U(1)$ symmetry spontaneously broken by the vevs $v_{1,2}$ of the two fields as well as by soft-breaking terms (see \cite{Barger:2008jx} for more details). Only the real scalar singlet~$S_{2}$ couples to the two heavy vector-like quarks~$\mathcal{B},\mathcal{T}$ which are charged under~$SU(3)_{c} \times U(1)_{\rm em}$ like their SM siblings. The Lagrangian of the scalar sector reads
\begin{align}
    -\mathcal{L} &\supset \mu_{\Phi}^{2} \vert \Phi \vert^{2} + \lambda_{\Phi} \vert \Phi \vert^{4} + \sum_{j} \mu_{j}^{2} \left\vert S_{j} \right\vert^{2} + \lambda_{j} \left\vert S_{j} \right\vert^{4} + \lambda_{\Phi j} \left\vert \Phi \right\vert^{2} \left\vert S_{j} \right\vert^{2} + \mu_{2j} S_{2} \vert S_{j} \vert^{2} \nonumber \\
    &\hspace{-5mm} + \mu_{\Phi 2} S_{2} \left\vert \Phi \right\vert^{2} + \lambda_{3} \left\vert S_{1} \right\vert^{2} \left\vert S_{2} \right\vert^{2} + \left( \tilde{\mu}_{\Phi 1} S_{1} \left\vert \Phi \right\vert^{2}  +\tilde{\mu}_{1} S_{1}^{3} + \sum_{j} \tilde{\mu}_{1j} S_{1} \left\vert S_{j} \right\vert^{2} + \mathrm{h.c.} \right)
\end{align}
with complex parameters~$\tilde{\mu}_{\Phi 1},\tilde{\mu}_{1},\tilde{\mu}_{1j}$, and the scalar multiplets
\begin{align}
    \Phi = \frac{1}{\sqrt{2}} \begin{pmatrix}
        \sqrt{2} G^{+} \\ v_{h} + h + i G^{0}
    \end{pmatrix} \quad , \quad S_{1} = \frac{1}{\sqrt{2}} \left( v_{1} + \hat{s}_{1} + i a_{1} \right) \quad , \quad S_{2} = v_{2} + \hat{s}_{2} \, .
\end{align}
The fields dressed with a hat are not the mass eigenstates yet. Note that the imaginary parts of the complex parameters, collectively defined as~$\tilde{\mu}$ here, lead to a mixing between $s_{j}$ and $a_{1}$. By restricting the parameters so that the component of the mass matrix responsible for mixing between the pseudoscalar and the scalars is much smaller than the (final) masses of the scalars in the theory, this mixing can be safely neglected, and the masses of the physical scalars and pseudoscalars are given by the eigenvalues of two separate and symmetric mass matrices. Additionally, for simplicity, we assume the following parameter relations within the mass matrices:~$\mathrm{Re}\tilde{\mu}_{\Phi 1} = - \lambda_{\Phi 1} v_{1}/\sqrt{2}$, $\mu_{\Phi 2} = -2 \lambda_{\Phi 2} v_{2}$, $\mathrm{Re} \tilde{\mu}_{1} = -\mathrm{Re} \tilde{\mu}_{11}$, $\mu_{21} = -2\lambda_{3} v_{2}$, and $\mu_{22} = -8\lambda_{2} v_{2}/3$, leading to no mass mixing between the Higgs doublet and the singlets. The mass matrices thus become
\begin{align}
    \mathcal{M}_{\mathrm{s}} &= \begin{pmatrix}
    2 \lambda_{\Phi} v_{h}^{2} & \times & \times \\
    0 & \frac{\lambda_{\Phi 1}}{2}v_{h}^{2} + 2\lambda_{1} v_{1}^{2} - \sqrt{2} \frac{v_{2}^{2}}{v_{1}} \mathrm{Re} \tilde{\mu}_{12} & \times \\
    0 & 2\sqrt{2}v_{2} \mathrm{Re} \mu_{12} & \lambda_{\Phi 2} v_{h}^{2} + \lambda_{3} v_{1}^{2}
    \end{pmatrix}\\
    \mathcal{M}_{\mathrm{ps}} &=\mathrm{diag} \left(m_{G_0}, m_{a_1}\right)= \mathrm{diag} \left( 0, \frac{\lambda_{\Phi 1}}{2}v_{h}^{2} - \sqrt{2} \frac{4v_{1}^{2} \mathrm{Re} \tilde{\mu}_{1} + v_{2}^{2} \mathrm{Re} \tilde{\mu}_{12}}{v_{1}} \right)\,.
    \label{eq:psmatrix_multi}
\end{align}
In addition to the two scalar singlets and the vector-like quarks~$\mathcal{B}$, $\mathcal{T}$, this model features a vector-like fermion~$\chi$ which transform as a singlet under the SM gauge group and as $\chi \rightarrow -\chi$ upon a discrete, stabilizing $\mathbb{Z}_{2}$ symmetry. Therefore, it serves as a dark matter candidate here. Assuming real Yukawa coupling parameters~$y_{S}, y_{\mathcal{Q}}$, the BSM fermion sector is described by
\begin{align}
    -\mathcal{L} \supset \mu_{\chi} \overline{\chi} \chi + \left( y_{S} \overline{\chi_{L}} S_{2} \chi_{R} + \mathrm{h.c.} \right) + \sum_{\mathcal{Q}=\mathcal{B},\mathcal{T}} \mu_{\mathcal{Q}} \overline{\mathcal{Q}} \mathcal{Q} + \left( y_{\mathcal{Q}}\overline{\mathcal{Q}_{L}} S_{2} \mathcal{Q}_{R} + \mathrm{h.c.} \right) \, ,
    \label{Eq:MultiSingletModel_FermionLagrangian}
\end{align}
leading to the BSM fermion masses~$m_{\chi} = \mu_{\chi} + y_{S} v_{2}$ and $m_{\mathcal{Q}} = \mu_{\mathcal{Q}} + y_{\mathcal{Q}} v_{2}$.

The mixing between the two BSM scalars is described by
\begin{align}
\begin{pmatrix}
    s_{1} \\ s_{2}        
    \end{pmatrix} = \mathcal{R}_{\theta} \begin{pmatrix}
    \hat{s}_{1} \\ \hat{s}_{2}     
    \end{pmatrix} \equiv \begin{pmatrix}
        \cos\theta & \sin\theta \\
        -\sin\theta & \cos\theta
    \end{pmatrix} \begin{pmatrix}
    \hat{s}_{1} \\ \hat{s}_{2}     
    \end{pmatrix} \, .
    \label{Eq:TwoScalarMixing}
\end{align}
Upon the considered assumptions, the set of free parameters of the model is the following:
\begin{align}
    &\{ v_{1,2}, m_{s_{1,2}}, m_{a_{1}}, m_{\chi}, m_{\mathcal{B},\mathcal{T}}, \mu_{\Phi}^{2}, \mu_{1,2}^{2}, \sin\theta,  y_{B,T}, y_{S}, \varphi_{\tilde{\mu}_{1}}, \varphi_{\tilde{\mu}_{11}}, \varphi_{\tilde{\mu}_{12}}, \varphi_{\tilde{\mu}_{\Phi 1}} \}
\end{align}
with~$\varphi_{\tilde{\mu}_{i}}$ being the complex phases of the parameters~$\tilde{\mu}_{i}$. 
In summary, once the physical mass basis is considered, three mediators are present in the model, the two scalars $s_{1,2}$ and the pseudoscalar $a_{1}$. While not the only possible choice, we will compare the multi-singlet model with the \eDMEFT variant featuring two CP-even scalar mediators. To this end, we integrate out the pseudoscalar $a_1$ at the scale $\Lambda$, using a parameter configuration that fulfills the most relevant theoretical and experimental constraints on the UV model, as summarized in Appendix~\ref{App:TwoSingletExtension_Constraints}, and allows masses of the pseudoscalar in Eq.~(\ref{eq:psmatrix_multi}) consistent with our EFT approach. 

The contributions to the operators in the last two lines of Eq.~(\ref{Eq:eDMEFT_Lagrangian}) arise at loop level and read~\cite{Altmannshofer:2015xfo}
\begin{align}
c_{G}^{\hat{s}_{1}}&= -\sum_{\mathcal{Q}} \frac{y_{\mathcal{Q}} s_{\theta} }{2\,m_{\mathcal{Q}}} F_{S}\left( \frac{m_{s_1}^{2}}{4 m_{\mathcal{Q}}^{2}} \right),\quad \quad  c_{B}^{\hat{s}_{1}}  =-\sum_{\mathcal{Q}} \frac{3\,Y_\mathcal{Q}^2\,y_{\mathcal{Q}} s_{\theta} }{m_{\mathcal{Q}}} F_{S}\left( \frac{m_{s_1}^{2}}{4 m_{\mathcal{Q}}^{2}} \right),\\ 
     c_{G}^{\hat{s}_{2}} &=  -\sum_{\mathcal{Q}} \frac{\,y_{\mathcal{Q}} c_{\theta}}{2\,m_{\mathcal{Q}}} F_{S}\left( \frac{m_{s_2}^{2}}{4 m_{\mathcal{Q}}^{2}} \right),\quad \quad c_{B}^{\hat{s}_{2}}=-\sum_{\mathcal{Q}} \frac{3\,Y_\mathcal{Q}^2\,y_{\mathcal{Q}} c_{\theta}}{m_{\mathcal{Q}}} F_{S}\left( \frac{m_{s_2}^{2}}{4 m_{\mathcal{Q}}^{2}} \right),
\end{align}
and $c_{W}^{\hat{s}_{1}} = c_{W}^{\hat{s}_{2}} = 0$ with the BSM fermions~$\mathcal{Q}=\mathcal{B},\mathcal{T}$ and their respective masses~$m_{\mathcal{Q}}$ and weak hypercharges~$Y_{\mathcal{Q}}$. 
The scalar loop functions $F_{S}$ are the same as in the previous section in Eq.~(\ref{Eq:MonoSingletModel_LoopFunctions}). The comparison of the present multi-singlet model and the \eDMEFT will be based on a scan over the following parameter ranges (remember that $m_Q \equiv \Lambda$):
\begin{align}
    &m_{s_{1},s_{2}} \in \left[ 2 m_{h}, \Lambda \right] \quad , \quad m_{\chi} \in \left[ 50\GeV, \Lambda \right] \quad , \quad v_{1,2} \in \left[ 100\GeV, 500\GeV \right] \quad , \nonumber \\
    &y_{S}, y_{\mathcal{Q}} \in \left[ -2,2 \right] \quad , \quad \mu_{\Phi,1,2}^{2} \in \left[ - \left( 500\GeV \right)^{2} , -\left( 10 \GeV \right)^{2} \right] \quad , \nonumber \\
    &\sin\theta \in \left[ 0,0.7 \right] \quad , \quad \arg \tilde{\mu}_{\Phi 1} \in \left[ 0,0.01 \right] \quad , \quad \arg \tilde\mu_{1}, \arg \tilde{\mu}_{11}, \arg \tilde{\mu}_{12} = 0 \, .
\end{align}

\subsubsection{Dark Matter relic abundance}

The comparison between the \eDMEFT and the multi-singlet model will proceed along the same lines as in the previous section. In the chosen setup, the DM relic abundance is governed by the same kind of processes as for the mono-singlet model.
\begin{figure}[b!]
    \centering
    \includegraphics{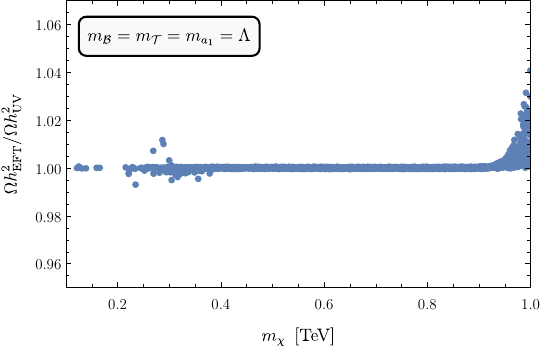}
    \caption{Ratio of the DM relic abundance for the \eDMEFT and the two-singlet extension with the new-physics scale~$\Lambda = 1\TeV$. The scalars~$s_{1}$, $s_{2}$ correspond to the light mediators in the \eDMEFT. }
    \label{Fig:MultiSingletModel_RatioRelicAbundance}
\end{figure}
To this end, Fig.~\ref{Fig:MultiSingletModel_RatioRelicAbundance} shows the ratio of the predictions for the DM relic density in the \eDMEFT and the concrete model. We point out that this ratio basically coincides with~$1$, unless the values of the DM mass approaches the cut-off scale~$\Lambda$.\footnote{The tiny deviation for DM masses $200\GeV \lesssim m_{\chi} \lesssim 400\GeV$, partially present already in Fig.~\ref{Fig:MonoSingletModel_RatioRelicAbundance}, is caused most likely by numerical inaccuracies.} This is expected since, analogously to the scenario discussed in the previous section, for $m_\chi \ll \Lambda$ the annihilation process is governed by momenta well within the validity range of the EFT, while for $m_\chi$ approaching the cutoff, the EFT loses validity since missing the pseudoscalar state $a_{1}$ in the spectrum. This consequently leads to $\Omega h^{2}_{\mathrm{UV}}< \Omega h^{2}_{\mathrm{EFT}}$ for masses very close to the cutoff.

\subsubsection{Dark Matter Direct Detection}
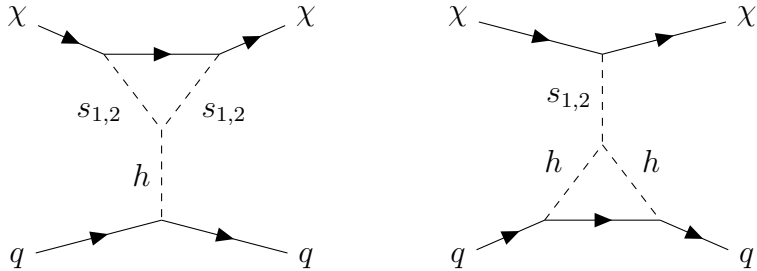
\begin{figure}[b!]
\centering
\begin{tikzpicture}
\begin{feynman}
\vertex (a1) {\( \chi \)};
\vertex[below=3.2cm of a1] (c1) {\( q \)};

\vertex[right=3.8cm of a1] (a4) {\( \chi \)};
\vertex[below=3.2cm of a4] (c3) {\( q \)};

\vertex at ($(c1)!0.5!(c3) + (0,0.5cm)$) (c2);

\vertex at ($(a1)!0.3!(a4) - (0,0.5cm)$) (a2);
\vertex at ($(a1)!0.7!(a4) - (0,0.5cm)$) (a3);

\vertex at ($(a2)!0.5!(a3) - (0,1cm)$) (b1);

\vertex[right=2cm of a4] (d1) {\( \chi \)};
\vertex[below=3.2cm of d1] (f1) {\( q \)};

\vertex[right=3.8cm of d1] (d3) {\( \chi \)};
\vertex[below=3.2cm of d3] (f4) {\( q \)};

\vertex at ($(d1)!0.5!(d3) - (0,0.5cm)$) (d2);

\vertex at ($(f1)!0.3!(f4) + (0,0.5cm)$) (f2);
\vertex at ($(f1)!0.7!(f4) + (0,0.5cm)$) (f3);

\vertex at ($(f2)!0.5!(f3) + (0,1cm)$) (e1);

\diagram* {
(a1) -- [fermion] (a2) -- [fermion] (a3) -- [fermion] (a4),
(c1) -- [fermion] (c2) -- [fermion] (c3),

(a2) -- [scalar, edge label'=\( s_{1,2} \)] (b1),
(a3) -- [scalar, edge label=\( s_{1,2} \)] (b1),

(b1) -- [scalar, edge label'=\( h \)] (c2),

(f1) -- [fermion] (f2) -- [fermion] (f3) -- [fermion] (f4),
(d1) -- [fermion] (d2) -- [fermion] (d3),

(d2) -- [scalar, edge label'=\( s_{1,2} \)] (e1),
(f2) -- [scalar, edge label=\( h \)] (e1),

(e1) -- [scalar, edge label=\( h \)] (f3),
};
\end{feynman}
\end{tikzpicture}
\caption{Generally possible one-loop scattering processes for DMDD in the multi-singlet model.}
\label{Fig:MultiSingletModel_DMDD_FeynmanDiagrams}
\end{figure}

Next, we investigate the model in light of DMDD. The effective Lagrangian has the same form of the mono-singlet model previously discussed, see Eq.~(\ref{Eq:DMDDeffectiveLagrangian}). However, since the mixing between the Higgs boson and the other CP-even scalars is neglected, the Wilson coefficients arise at the loop level. The coefficient $C_q$ originates from triangular topologies, as the ones shown in Fig.~\ref{Fig:MultiSingletModel_DMDD_FeynmanDiagrams} involving trilinear couplings between the Higgs and the BSM mediators:
\begin{align}
    C_{q} = C_{q}^{\mathrm{trig}}= \frac{1}{8 \pi^{2} v_{h} m_{h}^{2} m_{\chi}} \left[ g_{h s_{1} s_{2}} y_{s_{1}}y_{s_{2}} \mathcal{F} \left( \frac{m_{s_{1}}^{2}}{m_{\chi}^{2}},\frac{m_{s_{2}}^{2}}{m_{\chi}^{2}} \right) + \sum_{j=1}^{2} y_{s_{j}}^{2} g_{h s_{j} s_{j}} \mathcal{G} \left( \frac{m_{s_{j}}^{2}}{m_{\chi}} \right)  \right] \, ,
\end{align}
where the $g_{h s_{i} s_{j}}$ couplings are given in Eqs.~(\ref{eq:multi_cop_s1s1})-(\ref{eq:multi_cop_s1s2}) and 
\begin{align}
    y_{s_{1}} = y_{S} \sin \theta\quad , \quad y_{s_{2}} = y_{S} \cos \theta
\end{align}
with the BSM Yukawa coupling parameter~$y_{S}$ between the scalar~$S_{2}$ and the DM fermion~$\chi$ in Eq.~(\ref{Eq:MultiSingletModel_FermionLagrangian}).
Note that the second diagram in Fig.~\ref{Fig:MultiSingletModel_DMDD_FeynmanDiagrams} does not contribute in our specific scenario as the choice of model parameters renders the trilinear $s_{1,2}hh$~vertex zero. Defining the auxiliary function
\begin{align}
    f \left( x,y \right) &= \frac{(x-6) x  \log x}{2
   (x-y)}-\frac{(x-4)^{3/2} \sqrt{x}   \log \left[ \frac{1}{2} \left(\sqrt{x-4}+\sqrt{x}\right)\right]}{x-y} \, ,
\end{align}
the loop functions are given by
\begin{align}
\mathcal{F} \left( x,y \right) &= f\left( x,y \right) + f \left( y,x \right) - 1 \\
\mathcal{G} \left( x \right) &= (1-x) \sqrt{\frac{x-4}{x}}  \log \left[\frac{1}{2} \left(\sqrt{x-4}+\sqrt{x}\right)\right]  + \frac{1}{2} (x-3) \log x - 1\, .
\end{align}
The coefficient $C_{G}$ is the combination of two kinds of contributions, i.e.
\begin{align}
C_{G} = \sum_{Q=c,b,t}\frac{2}{27}C_{Q}^{\mathrm{trig}} + C_{G}^{\mathrm{VL}}
\end{align}
with $C_{Q}^{\mathrm{tri}}$ coming from the same triangle diagrams discussed before, while $C_{G}^{\mathrm{VL}}$ corresponds to an effective coupling between scalar bosons and gluons generated by a triangle loop of vector-like fermions. The latter is written as
\begin{align}
C_{G}^{\mathrm{VL}} = \frac{2\sqrt{2}}{9 m_{h}^{2}} \left( c_{G}^{\hat{s}_{1}} + c_{G}^{\hat{s}_{2}} + c_{G}^{h} \right) \, ,
\end{align}
where
\begin{align}
c_{G}^{\phi} = \frac{2 \sqrt{2} m_{\mathcal{Q}}^{2} \left( \left( m_{\phi}^{2} - 4 m_{\mathcal{Q}}^{2} \right) \arcsin \left( \frac{m_{\phi}}{2 m_{\mathcal{Q}}} \right)^{2} + m_{\phi}^{2} \right)}{m_{\phi}^{4} v_{\phi}} \, .
\end{align}
By combining everything one can define the SI scattering cross section of the DM fermion off nucleons, as done in Eq.~(\ref{Eq:DMDDSIcrossSection}), with the dimensionless parameter
\begin{align}
C_{N} = m_{N}\left[ \sum_{q=u, d, s} C_{q} f_{T_{q}}^{N} + C_{G} f_{T_{G}}^{N} \right] \, .
\end{align}
We have verified numerically that the Wilson coefficients described above can be perfectly matched with their counterparts in an \eDMEFT realization with two CP-even mediators as long as the propagating degrees of freedom are not close to the cut-off scale $\Lambda$.

\subsubsection{Collider signatures}
\begin{figure}[b!]
    \begin{tikzpicture}[baseline=7mm]
        \begin{feynman}
            \vertex (a1) {\(g\)};
            \vertex[above=1.5cm of a1] (a2) {\(g\)};

            \vertex[right=1cm of a1] (b1);
            \vertex[right=1cm of a2] (b2);

            \vertex at ($(b1)!0.5!(b2) + (0.7cm,0)$) (c);
            \vertex[right=1cm of c] (d);

            \vertex at ($(a1) + (3.7cm,0)$) (e1) {\(h\)};
            \vertex at ($(a2) + (3.7cm,0)$) (e2) {\(\mathcal{S}_{j}\)};

            \diagram* {
                (a1) -- [gluon] (b1),
                (a2) -- [gluon] (b2),
                (b1) -- [fermion,edge label=\(\mathcal{Q}\)] (b2),
                (b2) -- [fermion] (c),
                (c) -- [fermion] (b1),
                (c) -- [scalar, edge label=\(\mathcal{S}_{i}\)] (d),
                (e1) -- [scalar] (d),
                (e2) -- [scalar] (d)
            };
        \end{feynman}
    \end{tikzpicture}
    \hspace{2mm}
    \centering
    \begin{tikzpicture}[baseline=7mm]
        \begin{feynman}
            \vertex (a1) {\(g\)};
            \vertex[above=1.5cm of a1] (a2) {\(g\)};

            \vertex[right=1cm of a1] (b1);
            \vertex[right=1cm of a2] (b2);

            \vertex at ($(b1)!0.5!(b2) + (0.7cm,0)$) (c);
            \node[right=1cm of c,dot] (d);

            \vertex at ($(a1) + (3.7cm,0)$) (e1) {\(Z\)};
            \vertex at ($(a2) + (3.7cm,0)$) (e2) {\(S_i\)};

            \diagram* {
                (a1) -- [gluon] (b1),
                (a2) -- [gluon] (b2),
                (b1) -- [fermion,edge label=\(q\)] (b2),
                (b2) -- [fermion] (c),
                (c) -- [fermion] (b1),
                (c) -- [boson, edge label=\(Z/\gamma\)] (d),
                (e1) -- [boson] (d)  -- [scalar](e2)
            };
        \end{feynman}

    \end{tikzpicture}
    \hspace{2mm}
    \centering
    \begin{tikzpicture}[baseline=7mm]
        \begin{feynman}
            \vertex (j1) {\(\overline{q}\)};
            \vertex[above=1.5cm of j1] (j2) {\(q\)};

            \vertex at ($(j1)!0.5!(j2) + (1cm,0)$) (k);
            \node[right=1cm of k,dot] (l);

            \vertex at ($(j1) + (3cm,0)$) (m1) {\(Z\)};
            \vertex at ($(j2) + (3cm,0)$) (m2) {\(\mathcal{S}_{i}\)};
        
            \diagram* {
            (k) -- [fermion] (j1),
            (j2) -- [fermion] (k),
            (k) -- [boson,edge label=\(Z/\gamma\)] (l),
            (l) -- [boson] (m1),
            (m2) -- [scalar] (l),
            };
        \end{feynman}
    \end{tikzpicture}
    \caption{Processes contributing to mono-$h$ (first diagram) and mono-$Z$ signatures, with the coupling $S_i - Z/\gamma - Z$ arising at loop level due to the Wilson coefficient $c_B^{S_i}$. The mediators are $\mathcal{S}_{i,j} = s_1, s_2$.
    } 
    \label{Fig:MultiSingletModel_ColliderSignatures}
\end{figure}
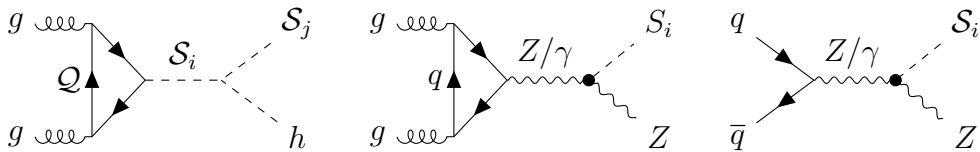
\begin{figure}[b!]
    \centering
    \includegraphics[width=0.49\textwidth]{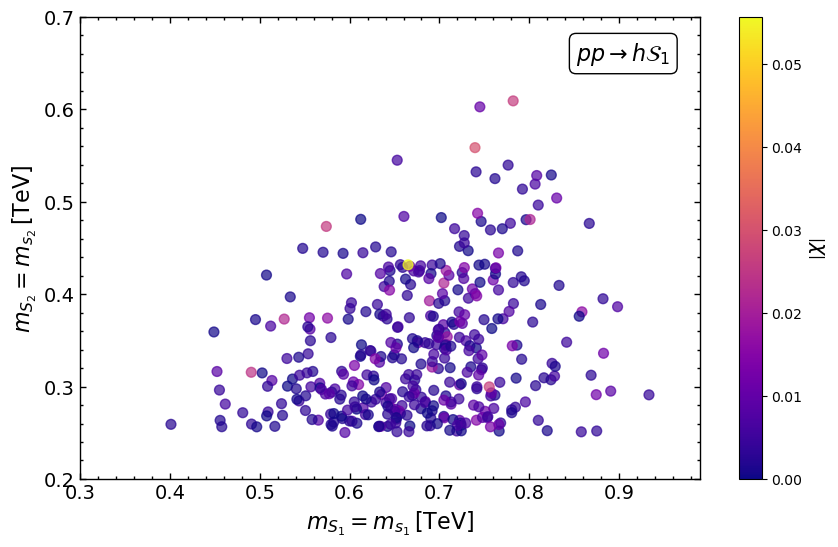}
    \includegraphics[width=0.49\textwidth]{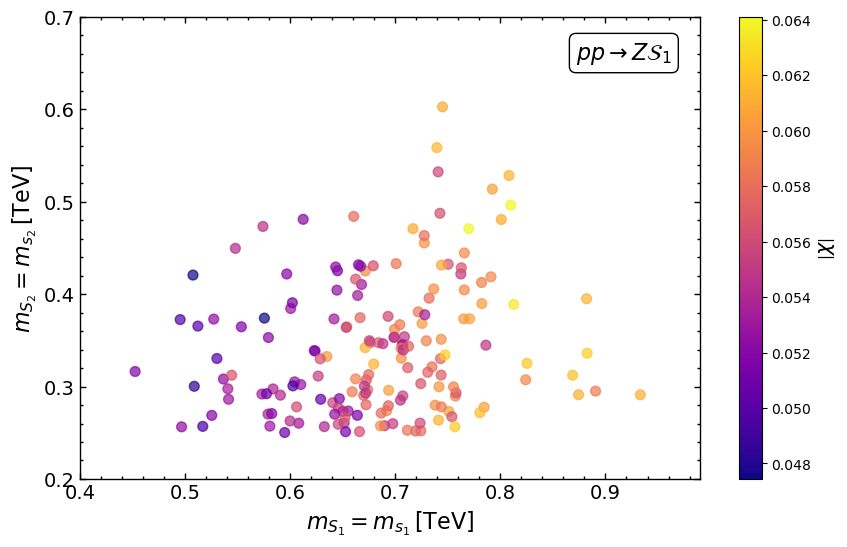}
    \caption{The mono-$h$~(\emph{left}) and mono-$Z$~(\emph{right}) production cross sections in the multi-singlet model are found to be in good agreement with the \eDMEFT, as indicated by the small relative difference $|\chi|\lesssim 5\%$.
    }
    \label{Fig:MultiSingletModel_Crossx}
\end{figure}

As the ultimate aspect of our investigation for this model, we shall discuss collider signatures, with the relevant diagrams depicted in~Fig.~\ref{Fig:MultiSingletModel_ColliderSignatures}. Since we are considering, for the comparison with the multi-singlet model an EFT with only CP-even mediators, the mono-$h$ is the dominant signature, since the mono-$Z$ signature is dominated by loop interactions. Nevertheless, both signals are compared with the EFT. The result of the parameter scan is presented in the $m_{\mathcal{S}_{1,2}}$~bidimensional plane in
Fig.~\ref{Fig:MultiSingletModel_Crossx}. 
In an analogous fashion like in the previous section, we have verified the capability of the \eDMEFT of reproducing the collider phenomenology of the multi-singlet model via a parameter scan over the ranges previously illustrated. The different colors correspond again to the parameter~$\chi$, i.e. the relative difference between the mono-$h$ production cross sections predicted by the \eDMEFT and multi-singlet models. Evidently, the small value of~$\chi$ indicates a good agreement between the EFT and the concrete model. The reason is the same as in the previous section: the mono-$h$ production processes are described by the same Feynman diagrams in both the \eDMEFT and the multi-singlet model. 

\subsection{2HDM + one singlet}

As the last case studied in this work, we compare the \eDMEFT~framework to models featuring a two-doublet Higgs sector, alongside an additional scalar $SU(2)$~singlet. For definiteness we will consider the variant of the model in which the scalar singlet is CP-odd. Before presenting the results, we will briefly review the most salient features of the 2HDM in the next subsection.

\subsubsection{Introduction to the 2HDM}\label{Sec:2HDM}

Let us first discuss the two-doublet sector of the theory. The most general Lagrangian describing two $SU(2)$ Higgs doublets interacting with the SM reads~\cite{Branco:2011iw}
\begin{align}
    \mathcal{L}_{\mathrm{2HDM}} = \, & \sum_{j=1,2}\left\vert D_{\mu} \Phi_{j} \right\vert^{2} - y_{j,mn}^{d} \overline{Q}_{m,L} \Phi_{j} d_{n,R} - y_{j,mn}^{u} \overline{Q}_{m,L} \tilde{\Phi}_{j} u_{n,R} - y_{j,mn}^{e} \overline{L}_{m,L} \Phi_{j} e_{n,R} \nonumber \\
    &-M_{11}^{2} \left\vert \Phi_{1} \right\vert^{2} - M_{22}^{2} \left\vert \Phi_{2} \right\vert^{2} - M_{12}^{2} \left( \Phi_{2}^{\dagger}\Phi_{1} + \mathrm{h.c.} \right) - \frac{\lambda_{1}}{2} \left\vert \Phi_{1} \right\vert^{4} - \frac{\lambda_{2}}{2} \left\vert \Phi_{2} \right\vert^{4} \nonumber \\
    &- \lambda_{3} \left\vert \Phi_{1} \right\vert^{2} \left\vert \Phi_{2} \right\vert^{2} - \lambda_{4} \left\vert \Phi_{1}^{\dagger} \Phi_{2} \right\vert^{2} - \frac{1}{2} \lambda_{5} \left[  \left( \Phi_{2}^{\dagger}\Phi_{1} \right)^{2} + \mathrm{h.c.} \right] \nonumber \\
    &- \left( \lambda_{6} \left\vert \Phi_{1} \right\vert^{2} + \lambda_{7} \left\vert \Phi_{2} \right\vert^{2} \right) \left( \Phi_{2}^{\dagger}\Phi_{1} + \mathrm{h.c.} \right)\,,
    \label{Eq:2HDM_Lag}
\end{align}
with the gauge covariant derivative~$D_{\mu}$. The Higgs doublets can be decomposed as
\begin{align}
    \Phi_{j} = \frac{1}{\sqrt{2}} \begin{pmatrix}
        \sqrt{2}\hat{\phi}_{j}^{+} \\ v_{j} + \hat{\rho}_{j} + i \hat{\eta}_{j}
    \end{pmatrix} \quad \mathrm{with} \ j=1,2 \, , \label{Eq:HiggsDoubletsDef}
\end{align}
and we used the shorthand notation $\tilde{\Phi}_{j} \equiv i\sigma^{2}\Phi_{j}^{*}$. Fields dressed with a hat are not mass eigenstates and their rotations to the mass basis, similar to the ones in the previous section, will be discussed below. Rather than working with the general Lagrangian in Eq.~(\ref{Eq:2HDM_Lag}), it is customary to impose a softly broken discrete $\mathbb{Z}_{2}$~symmetry with the doublets transforming as $\Phi_{1} \mapsto \Phi_{1}$ and $\Phi_{2} \mapsto -\Phi_{2}$, so that $\lambda_{6,7} = 0$ while $M_{12}^{2}\neq 0$ in general. Furthermore, the scalar potential is made CP conserving by taking the parameters~$M_{12}^{2}$ and $\lambda_{5}$ to be real. To perform the matching with the \eDMEFT, it is useful to introduce the so called Higgs basis wherein the two Higgs doublets are defined by
\begin{align}
\label{eq:rot}
    \begin{pmatrix}
    \Phi_{h} \\ \Phi_{H}        
    \end{pmatrix} = \mathcal{R}_{\beta} \begin{pmatrix}
    \Phi_{1} \\ \Phi_{2}     
    \end{pmatrix} \, ,
\end{align}
with the rotation matrix~$\mathcal{R}_{\beta}$ introduced in Eq.~(\ref{Eq:TwoScalarMixing}), and the angle required for this is given by
\begin{align}
    \tan \beta = \frac{v_{2}}{v_{1}} \quad \mathrm{with} \ v_{1}^{2} + v_{2}^{2} = v_{h}^{2} \equiv \left( 246\GeV \right)^{2} \, .
\end{align}
The characteristic feature of the Higgs basis is the fact that only one of the two doublets acquires a vev. Indeed, the two doublets are decomposed as
\begin{align}
    \Phi_{h} = \frac{1}{\sqrt{2}} \begin{pmatrix}
        \sqrt{2}G^{+} \\ v_{h} + \hat{h} + i G^{0}
    \end{pmatrix} \quad , \quad \Phi_{H} = \frac{1}{\sqrt{2}} \begin{pmatrix}
        \sqrt{2}H^{+} \\ \hat{H} + i A
    \end{pmatrix} \, .
\end{align}
The scalars $G^{\pm}$, $G^{0}$ are the would-be Goldstone bosons and give rise to the longitudinal polarizations of the massive EW gauge bosons $W^{\pm}$, $Z$ after electroweak symmetry breaking. On the other hand, $A$ and $H^\pm$ are, respectively, pseudoscalar and electrically charged BSM Higgs bosons. Lastly, the weak eigenstates $\hat{h}$ and $\hat{H}$ are linear combinations of the $125\GeV$~Higgs boson and a further CP-even physical state. In order to avoid constraints from the Higgs signal strengths, see e.g.~\cite{ATLAS:2021vrm} for the most recent results, the so-called Higgs alignment limit will be imposed which is a mathematical condition ensuring that no mixing occurs between the CP-even scalars. With that we have~$\hat{h}\equiv h$, identified as the $125\GeV$~SM-like Higgs boson, and $\hat{H}\equiv H$. 

When extending the scalar sector of the SM, the emergence of large flavor changing neutral currents~(FCNC) should be prevented. A necessary, but not sufficient condition to achieve this  is to take, at the tree-level, for the two Higgs doublets the same Yukawa breaking structure as in the SM. In this setup, the Yukawa Lagrangian for the electrically neutral physical states can be written as:
\begin{align}
\mathcal{L}_{\mathrm{Yukawa}}=-\sum_f \left[y_{H}^{f}  H \overline{f}f + y_{h}^{f}  h \overline{f} f-i y_{H}^{f}\sin\theta a \overline{f}\gamma_5 f-i y_{H}^{f}\cos\theta A \overline{f}\gamma_5 f\right],
\end{align}
where the couplings can be expressed as the product between the SM Yukawa and a scaling parameter:
\begin{align}
    y_{\phi}^{f} = \epsilon_{f}^{\phi} y_{\mathrm{SM}}^{f} \quad \quad \mathrm{with} \quad y_{\mathrm{SM}}^{f} \equiv \sqrt{2}\frac{m_{f}}{v_{h}}
\end{align}
for the fermions~$f=u,d,e$ (where $u$ and $d$ represent the six up- and down-type quarks and $e$ the three electrically charged leptons) with respective masses~$m_{f}$ and~$\phi=H,h$. This leads to the so called Aligned Yukawa Model~\cite{Pich:2009sp,Tuzon:2010vt,Pich:2010ic,Penuelas:2017ikk,Gori:2017qwg}. Deviations from the SM Yukawa structure might, however, arise at the loop level. To avoid this possibility one often relies on the stronger assumption of the presence of additional $\mathbb{Z}_{2}$~symmetries, forcing the doublets to couple selectively to the fermionic fields.\footnote{The $\mathbb{Z}_{2}$ symmetry can be superseded, in a further extended framework, by a gauge symmetry. See e.g.~\cite{Campos:2017dgc,Arcadi:2020aot}.} There are four possible ways to impose such symmetries, leading to the Type-I, Type-II, Type-X (or lepton specific) and Type-Y (or flipped) variants of the 2HDM. In the alignment limit (i.e.~$\epsilon_{f}^{h} = 1$), the coupling modifiers of the BSM bosons are functions of~$\tan\beta$ and are summarized in Tab.~\ref{Tab:AlignedYukawaModel}, where we dropped the superscript $H$ for simplicity.
\begin{table}[b!]
    \centering
    \begin{tabular}{cccccc}
        \toprule
        \multicolumn{1}{c}{Modifier} & \multicolumn{1}{c}{Type I} & \multicolumn{1}{c}{Type II} & \multicolumn{1}{c}{Type X} & \multicolumn{1}{c}{Type Y} & \multicolumn{1}{c}{Inert} \\
        \midrule
        $\epsilon_{u}$ & $\cot \beta$ & $\cot \beta$ & $\cot \beta$ & $\cot \beta$ & $0$ \\
        $\epsilon_{d}$ & $\cot \beta$ & $-\tan \beta$ & $\cot \beta$ & $-\tan \beta$ & $0$ \\
        $\epsilon_{e}$ & $\cot \beta$ & $-\tan \beta$ & $-\tan \beta$ & $\cot \beta$ & $0$ \\
        \bottomrule
    \end{tabular}
    \caption{Modifiers of Yukawa interactions between the BSM Higgs boson and fermions~\cite{Arcadi:2020gge}.}
    \label{Tab:AlignedYukawaModel}
\end{table}
In the following we will focus on the Type-II 2HDM, in which $\Phi_{2}$ couples to the up-type quarks while $\Phi_{1}$ couples to the down-type quarks and charged leptons, and further extend the scalar sector of the theory by different $SU(2)_{L}$~singlets.

\subsubsection{2HDM + pseudoscalar singlet + fermionic DM}
Assuming softly-broken CP symmetry in the Higgs sector, the extension of the 2HDM potential with an additional real pseudoscalar~$\hat{P}$ in the interaction basis it is commonly defined as~\cite{Arcadi:2020gge}
\begin{align}
    -\mathcal{L}_{P} \supset \frac{M_{PP}^{2}}{2} \hat{P}^{2} + \frac{\lambda_{P}}{4}\hat{P}^{4} + \lambda_{11P} \left\vert \Phi_{1} \right\vert^{2} \hat{P}^{2} + \lambda_{22P} \left\vert \Phi_{2} \right\vert^{2} \hat{P}^{2} + \left( i\mu_{12P} \Phi_{1}^{\dagger}\Phi_{2}\hat{P} + \mathrm{h.c.} \right)  \label{Eq:2HDM+PS_Lag}
\end{align}
and the Lagrangian for the fermionic DM is
\begin{align}
    \mathcal{L}_{\chi} = \overline{\chi}i\slashed{\partial} \chi - m_{\chi} \overline{\chi} \chi - i y_{\chi P}\overline{\chi} \gamma^{5} \chi \hat{P} \, .
\end{align}
After EW symmetry breaking, mass mixing between the singlet field~$\hat{P}$ and the pseudoscalar state~$\hat{A}$ present in the doublet sector occurs. The entries of the pseudoscalar mass matrix are given by
\begin{align}
    m_{AA}^{2} &= - \left( 2 M_{12}^{2}/s_{2\beta} + \lambda_{5} v_{h}^{2} \right) \\
    m_{AP}^{2} &= -\mu_{12P} v_{h} \\
    m_{PP}^{2} &= M_{PP}^{2} + \left( \lambda_{11P}c_{\beta}^{2} + \lambda_{22P} s_{\beta}^{2}\right) v_{h}^{2} \, .
\end{align}
The latter can be diagonalized via a rotation matrix $\mathcal{R}_\theta$ with the mixing angle given by 
\begin{align}
    \tan 2\theta &= \frac{2 m_{AP}^{2}}{m_{PP}^{2} - m_{AA}^{2}} \, .
\end{align}
The set of the physical states emerging from the Higgs sector is, in summary, composed of two CP-even states, $h,H$, two CP-odd states, $a,A$ and two electrically charged scalars~$H^{\pm}$. Their masses can be written in term of the parameters of the scalar potential as
\begin{align}
    m_{h}^{2} &= \left( \lambda_{2} \sin^{2} \beta + \lambda_{345} \cos^{2} \beta \right) v_h^{2} \\
    m_{H}^{2} &= -\frac{2M_{12}^{2}}{\sin 2\beta} + \left( \lambda_{2} - \lambda_{345} \right) v_h^{2} \sin^{2} \beta \\
    m_{H^{\pm}}^{2} &= -\frac{2M_{12}^{2}}{\sin 2\beta} - \frac{\lambda_{4} + \lambda_{5}}{2} v_h^{2} \\
    m_{A,a}^{2} &= \frac{1}{2\cos 2\theta} \left( \left( \cos 2\theta \pm 1 \right) m_{AA}^{2} + \left( \cos 2\theta \mp 1 \right) m_{PP}^{2} \right) \, .
\end{align}
As mentioned above, experimental results suggest interactions strengths between the scalar~$h$ and the other SM particles as described by the SM. Hence, this scalar is rendered SM-like here by imposing the alignment condition
\begin{align}
    \lambda_{1} &= \frac{\lambda_{2} \sin^{2} \beta + \lambda_{345} \cos 2\beta}{\cos^{2} \beta} \, .
\end{align}
Similarly to~\cite{Arcadi:2022lpp}, we will adopt for our numerical study the following set of free parameters:
\begin{align}
    \{ m_{H}^{2} \, , \, m_{A}^{2} \, , \, m_{a}^{2} \, , \, m_{H^{\pm}}^{2} \, , \, m_{\chi}   \, , \, \theta \, , \, \beta \, , \, \lambda_{3} \, , \, \lambda_{P} \, , \, \lambda_{11P} \, , \, \lambda_{22P} \, , \, y_{\chi P} \} \, .
\end{align}
\begin{figure}[b!]
    \centering
    \includegraphics[width=0.23\linewidth]{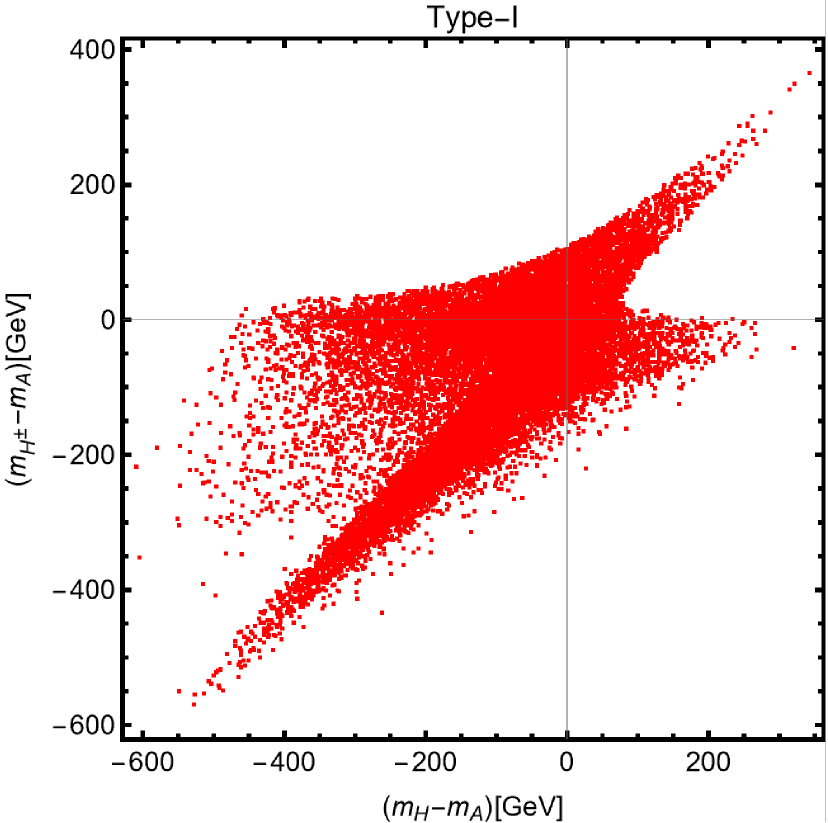}    \includegraphics[width=0.23\linewidth]{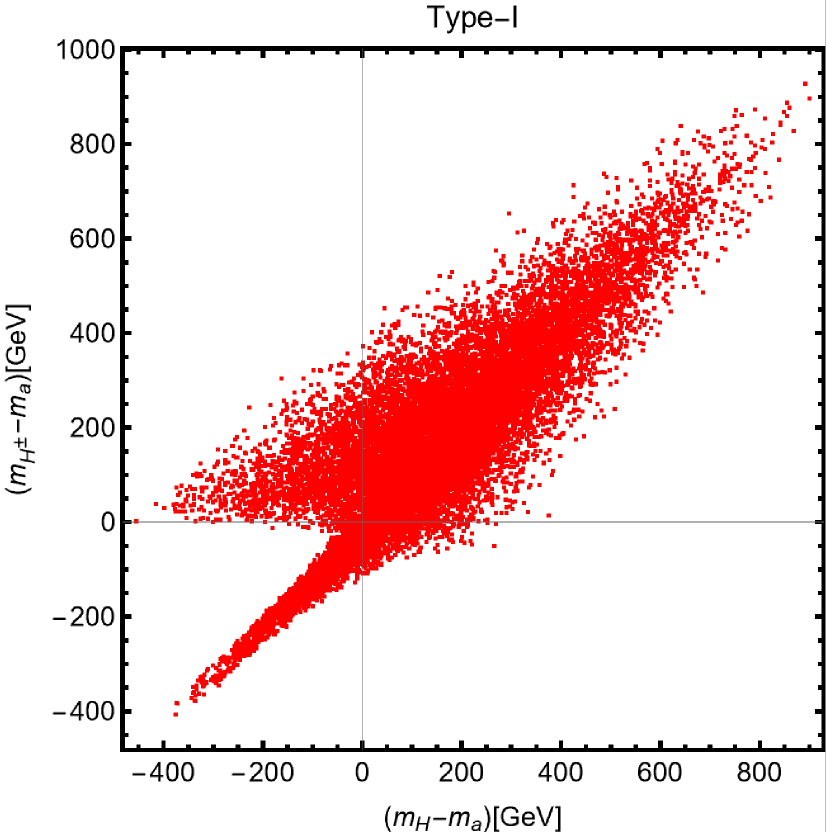}    \includegraphics[width=0.23\linewidth]{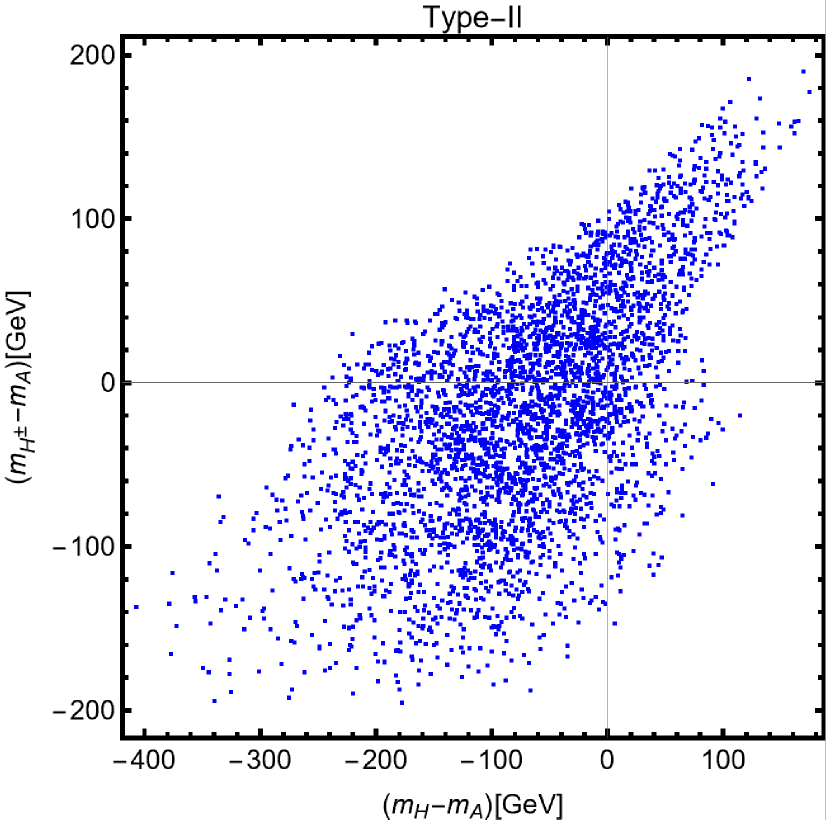}    \includegraphics[width=0.23\linewidth]{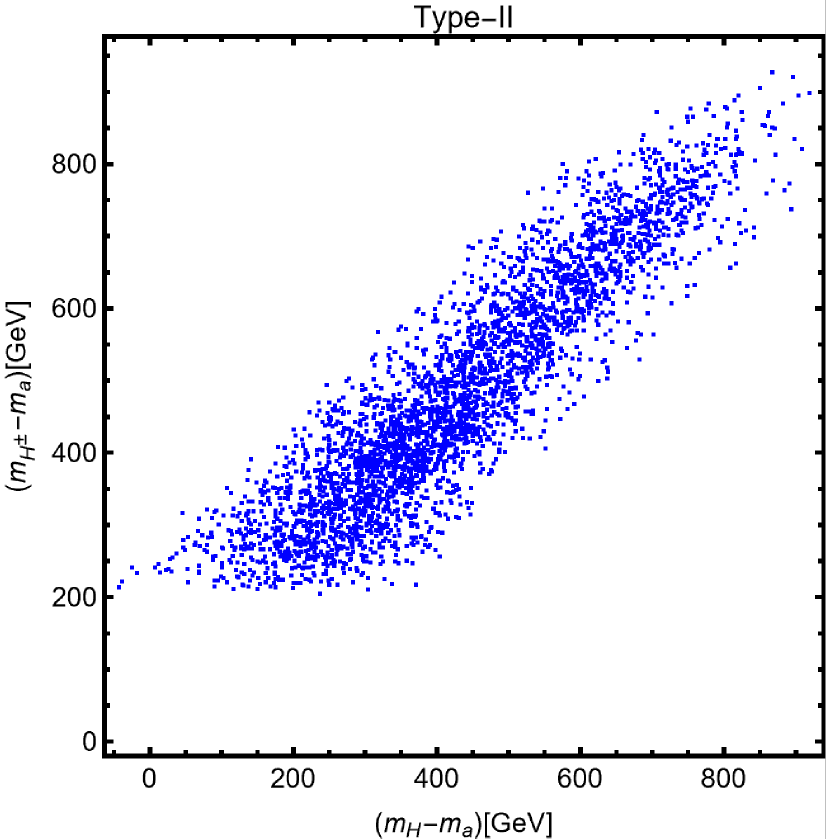}
    \caption{Mass splitting, in the 2HDM+a, between the $H$ and $H^\pm$ bosons and the two pseudoscalars.}
    \label{Fig:2HDMa_split}
\end{figure}
The 2HDM+a will be compared to two variants of the non-linear \eDMEFT framework, each covering different parameter space. First, we will consider the case in which both the $\mathcal{S}_{1,2}$ mediators are pseudoscalar and we hence identify $\mathcal{S}_1 \equiv a$, $\mathcal{S}_2 \equiv A$. The mass of the other BSM bosons of the 2HDM+a will be identified with the cut-off scale~$\Lambda$ of the \eDMEFT. In the second case, though, the CP-even state $H$, together with the lightest pseudoscalar $a$, will be identified with the propagating degrees of freedom of the \eDMEFT while the mass of the heavier pseudoscalar~$A$ and the charged Higgs bosons will be identified with the scale~$\Lambda$. As will be discussed in greater detail below, the accuracy of the EFT description of the 2HDM+a will depend on the extend to which it is possible to decouple, i.e. mass-split, the extra degrees of freedom, namely the charged Higgs $H^{\pm}$ and either $H$ or $A$, according to the chosen EFT realization. We note that, in order to integrate out parts of the second EW doublet, we write the doublet fields in polar coordinates with the Goldstone matrix $\Sigma$ parametrizing the phases of each. In consequence, the Higgs-like bosons transform as $SU(2)_L$ singlets~\cite{Lee:1972yfa,Weinberg:1978kz,Feruglio:1992wf,Contino:2010rs} (see also the recent~\cite{Buchalla:2023hqk}) and can straightforwardly be integrated out individually, leading to the {\it non-linear} EFT introduced in Section~\ref{Sec:IntroEFT}.

As evidenced in Fig.~\ref{Fig:2HDMa_split} the possibility of decoupling is limited from theoretical consistency of the scalar potential of the 2HDM+a, namely boundedness from below and perturbative unitarity, and EW precision tests~(EWPT), as detailed in Appendix~\ref{App:2HDMa_Constraints}. The figure illustrates the outcome the same parameter scan as performed in~\cite{Arcadi:2022lpp}. All the parameter assignations compatible with the aforementioned constraints are shown in the $(m_H-m_A,m_{H^\pm}-m_A)$ and $(m_H-m_a,m_{H^\pm}-m_a)$ plane. The red and blue points correspond, respectively to the Type-I and Type-II configurations of the Yukawa couplings of the BSM Higgs states. While the pseudoscalar $a$ can have a substantial mass splitting with the other states, allowing for a good EFT description, we observe that the bounds on the 2HDM+a, in particular EWPT, make it more challenging to achieve a sizable splitting between the $H,A,H^{\pm}$ states.

\begin{figure}[b!]
    \centering
    \includegraphics[width=0.49\textwidth]{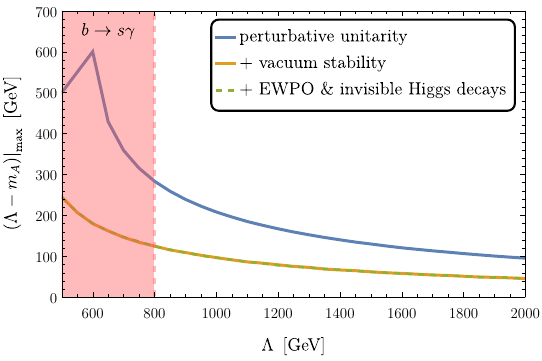}
    \includegraphics[width=0.49\textwidth]{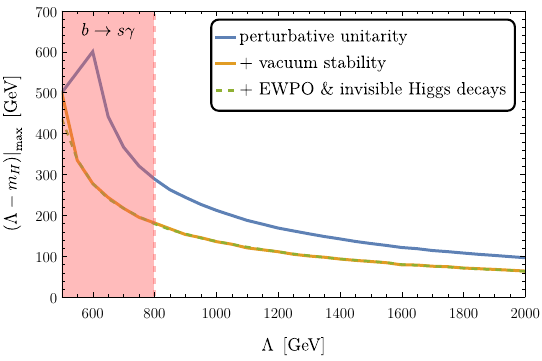}
    \caption{Maximum mass splitting between doublet scalars and the pseudoscalar $A$ (\emph{left}) and scalar $H$ (\emph{right}) for different sets of theoretical and experimental constraints. The electrically charged scalar resides at the new-physics scale~$\Lambda$ while in the left plot $m_{H} = \Lambda$ and in the right plot $m_{A} = \Lambda$. The mass of the singlet pseudoscalar is restricted to $m_{a} < m_{A}$. Measurements of $B$~meson branching ratios rule out $m_{H^{\pm}} < 800\GeV$ (see text for details).}
    \label{Fig:maxMassSplit_Aa}
\end{figure}

The dependence of the maximally possible mass splitting between the new-physics scale, commonly assumed as $m_{H^{\pm}}\equiv m_{H} \equiv \Lambda$, and the pseudoscalar~$A$ on the scale~$\Lambda$ is summarized in the left panel of Fig.~\ref{Fig:maxMassSplit_Aa}, whereas the right panel shows the analogous plot for the scalar $H$.
We inspect that, as expected, a strict decoupling of the degrees of freedom of the 2HDM+a not present in the EFT to which it is compared is not possible. As we will see below, it will nevertheless be possible that the EFT successfully describes the 2HDM+a in the phenomenologically relevant processes.
For the corresponding numerical scan of the EFT with two pseudoscalar mediators we will consider\begin{align}
    &m_{A} \in \left[ \Lambda - 100\GeV, \Lambda \right] \quad , \quad m_{H,H^{\pm}} = \Lambda \quad , \quad m_{a} \in \left[ 1\GeV, \Lambda \right] \nonumber \\
    &m_{\chi} \in \left[ m_{a}/10, \Lambda \right] \quad , \quad \lambda_{11P,22P} \in \left[ 0,4\pi \right] \quad , \quad \lambda_{3} \in \left[ 0.01, 4\pi \right] \nonumber \\
    &\beta \in \left[ \pi/4, 0.468\pi \right] \quad , \quad \theta \in \left[ 0,0.7 \right] \quad , \quad y_{\chi P} \in \left[ 0.01,3 \right] \, .
\end{align}
Following the reasoning above, we will always take $a$ as the lightest mediator and assume a relatively small splitting between the other pseudoscalar $A$ and the other BSM bosons residing at $\Lambda$. For the comparison between the 2HDM+a and the \eDMEFT with a scalar and a pseudoscalar mediator, we will just interchange the range of variation of $m_H$ and $m_A$.

\subsubsection{Dark Matter relic abundance}
\begin{figure}[b!]
    \centering
    \includegraphics[width=0.49\textwidth]{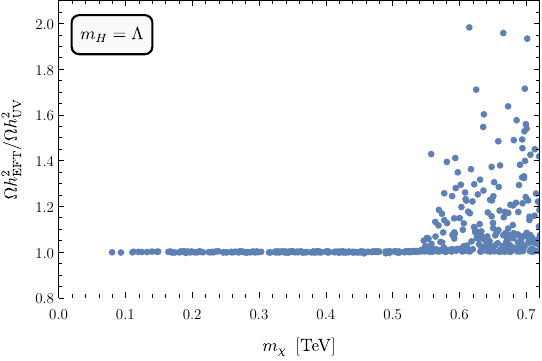}
    \includegraphics[width=0.49\textwidth]{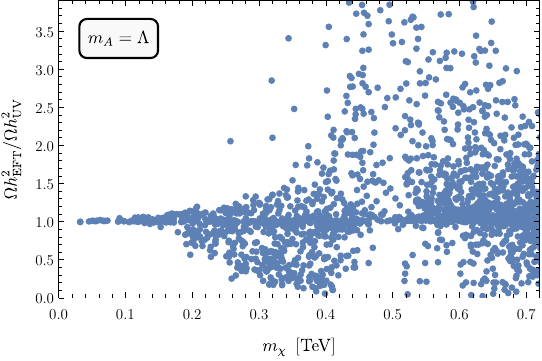}
    \caption{Ratio of the DM relic abundance for the \eDMEFT and the 2HDM+a with the new-physics scale~$\Lambda = 1\TeV$. \emph{Left}: The pseudoscalars~$A$, $a$ are the light mediators in the \eDMEFT. \emph{Right:} The fields $H$, $a$ act as mediators in the \eDMEFT.}
    \label{Fig:2HDMa_RatioRelicAbundance}
\end{figure}

To assess the capability of capturing the relevant features of the UV model with the \eDMEFT, we shall begin with the comparison of the DM relic abundance in each model. The results are shown in Fig.~\ref{Fig:2HDMa_RatioRelicAbundance}.
The left and right panels of the figure refer to, respectively, the case in which the heavy CP-even and the heavy CP-odd states are integrated out at the EFT cutoff scale~$\Lambda$. In the case that the mediators~$\mathcal{S}_{1,2}$ are both pseudoscalars, the \eDMEFT is very successful in reproducing the 2HDM+a up to DM masses around $m_\chi \lesssim 600\GeV \approx \Lambda/2$, as here the DM relic density is mostly due to annihilation processes into SM states via s-channel exchange of the $a/A$ states as well as annihilations into $aa,aA,AA$ final states via $t/u$-channel exchange of the DM. Discrepancies between the predictions of the EFT and the complete model only emerge for $m_{\chi} \gtrsim \Lambda/2$. In such a regime, new annihilation channels become kinematically accessible in the 2HDM+a, including states, namely $H$ and $H^{\pm}$, not present in the \eDMEFT. Consequently, the DM relic density in the 2HDM+a is smaller than in the EFT. Moving to the outcome of the right panel of Fig.~\ref{Fig:2HDMa_RatioRelicAbundance}, we notice differences in the predictions of the EFT and the UV model over a wider range of DM masses. A numerically accurate EFT description is essentially limited to $m_{\chi} \lesssim 0.2 \Lambda$. The reason is that in the EFT the DM is coupled only to one pseudoscalar mediator, while in the full model both can be exchanged in the s-channel. Consequently, for $2 m_{\chi} \gtrsim m_{A} \equiv \Lambda$, i.e.~when the $s$-channel exchange of the state~$A$ affects the DM annihilation into SM fermions in an unsuppressed way, the EFT breaks down, while already before the missing higher power corrections lead to deviations, as expected for any EFT. The starkest discrepancy between the relic abundance in the \eDMEFT and the one in the UV model at~$m_{\chi} \approx \Lambda/2$ is due to the $s$-channel resonance with the mediating heavy pseudoscalar~$A$. In fact, a cancellation between the two $s$-channel processes can emerge, resulting in an underestimation of the relic density in the EFT.
To summarize the results, the \eDMEFT captures well the DM physics in the regions where EFT validity is expected on general grounds.

\subsubsection{Dark Matter Direct Detection}\label{Sec:2HDM_DMDD}

\begin{figure}[b]
\centering
\begin{tikzpicture}
\begin{feynman}
\vertex (a1) {\( \chi \)};
\vertex[below=3.2cm of a1] (c1) {\( q \)};

\vertex[right=3.8cm of a1] (a4) {\( \chi \)};
\vertex[below=3.2cm of a4] (c3) {\( q \)};

\vertex at ($(c1)!0.5!(c3) + (0,0.5cm)$) (c2);

\vertex at ($(a1)!0.3!(a4) - (0,0.5cm)$) (a2);
\vertex at ($(a1)!0.7!(a4) - (0,0.5cm)$) (a3);

\vertex at ($(a2)!0.5!(a3) - (0,1cm)$) (b1);

\vertex[right=1.0cm of a4] (d1) {\( \chi \)};
\vertex[below=3.2cm of d1] (e1) {\( q \)};
\vertex[right=3.8cm of d1] (d4) {\( \chi \)};
\vertex[below=3.2cm of d4] (e4) {\( q \)};

\vertex at ($(d1)!0.3!(d4) - (0,0.5cm)$) (d2);
\vertex at ($(d1)!0.7!(d4) - (0,0.5cm)$) (d3);

\vertex at ($(e1)!0.3!(e4) + (0,0.5cm)$) (e2);
\vertex at ($(e1)!0.7!(e4) + (0,0.5cm)$) (e3);

\vertex[right=1.0cm of d4] (f1) {\( \chi \)};
\vertex[below=3.2cm of f1] (g1) {\( q \)};
\vertex[right=3.8cm of f1] (f4) {\( \chi \)};
\vertex[below=3.2cm of f4] (g4) {\( q \)};

\vertex at ($(f1)!0.3!(f4) - (0,0.5cm)$) (f2);
\vertex at ($(f1)!0.7!(f4) - (0,0.5cm)$) (f3);

\vertex at ($(g1)!0.3!(g4) + (0,0.5cm)$) (g2);
\vertex at ($(g1)!0.7!(g4) + (0,0.5cm)$) (g3);

\vertex at ($(f2)!0.5!(g3)$) (h) {};

\diagram* {
(a1) -- [fermion] (a2) -- [fermion] (a3) -- [fermion] (a4),
(c1) -- [fermion] (c2) -- [fermion] (c3),

(a2) -- [scalar, edge label'=\( A/a \)] (b1),
(a3) -- [scalar, edge label=\( A/a \)] (b1),

(
b1) -- [scalar, edge label'=\( h/H \)] (c2),

(d1) -- [fermion] (d2) -- [fermion] (d3) -- [fermion] (d4),
(e1) -- [fermion] (e2) -- [fermion] (e3) -- [fermion] (e4),

(d2) -- [scalar, edge label'=\( A/a \)] (e2),
(d3) -- [scalar, edge label=\( A/a \)] (e3),

(f1) -- [fermion] (f2) -- [fermion] (f3) -- [fermion] (f4),
(g1) -- [fermion] (g2) -- [fermion] (g3) -- [fermion] (g4),

(f2) -- [scalar] (g3),
(f3) -- [scalar] (h) -- [scalar] (g2)
};
\end{feynman}
\node[] at (10.7,-1.2) {$A/a$};
\node[] at (12.3,-1.2) {$A/a$};
\end{tikzpicture}
\caption{Relevant one-loop scattering processes for DMDD in the 2HDM+a model.}
\label{Fig:2HDMa_DMDD_FeynmanDiagrams}
\end{figure}
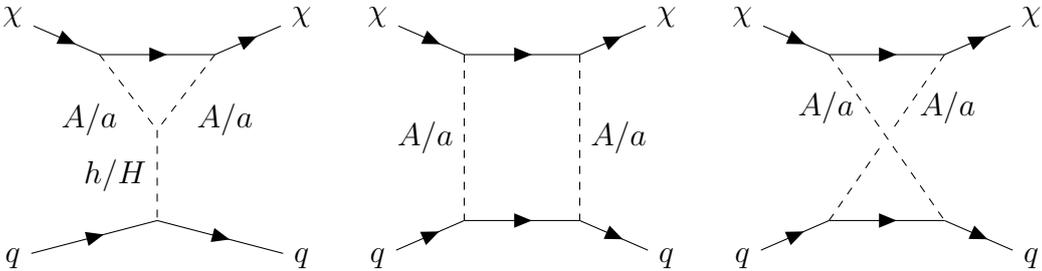

As discussed in~\cite{Dolan:2014ska,Arcadi:2024ukq} for instance, the most relevant operators for DMDD originate at the loop level in the case of fermionic DM interacting via a $s$-channel pseudoscalar portal.
The prominent topologies are illustrated in Fig.~\ref{Fig:2HDMa_DMDD_FeynmanDiagrams}.
Note that operators originating from tree-level interactions via $t$-channel exchange of the pseudoscalar mediator are not strictly absent, but largely subdominant with respect to the SI loop-induced interactions since they are, in the non-relativistic limit, momentum-suppressed and not coherent (as dependent on the DM and nucleon's spin.).

The effective Lagrangian describing the loop-induced SI interactions mostly responsible for DM scattering off nucleons is given by~\cite{Abe:2018emu}
\begin{align}
\mathcal{L}_{\mathrm{eff}}= & \frac{1}{2} \sum_{q} C_q m_q \overline{\chi} \chi \overline{q} q - \frac{9 \alpha_s}{16 \pi} C_{G} \overline{\chi} \chi G_{\mu \nu}^{a} G^{a \mu \nu} +\frac{1}{2} \sum_{Q} \overline{\chi} \left( C_{ Q}^{(1)} i \partial^\mu \gamma^{\nu} - C_{ Q}^{(2)} \partial^{\mu} \partial^{\nu} \right) \chi \mathcal{O}_{\mu \nu}^{ Q}
\label{Eq:DMDDeffectiveLagrangianWithTwist2Operator}
\end{align}
with the quarks~$q=u,d,s$ and $Q=u,d,s,c,b$ and the twist-2 operator
\begin{align}
\mathcal{O}_{\mu \nu}^{ Q}=\frac{i}{2} \overline{Q}\left(D_\mu \gamma_\nu+D_\nu \gamma_\mu-\frac{1}{2} g_{\mu \nu} \slashed{D} \right)  Q \, .
\end{align}
Each of the Wilson coefficients in the effective Lagrangian is the sum of contributions from diagrams with box-shaped and triangle-shaped topology. They can be decomposed as
\begin{align}
C_{q} = C_{q}^{\mathrm{tri}} + C_{q}^{\mathrm{box}} \quad , \quad C_{G} = C_{G}^{\mathrm{tri}} + C_{G}^{\mathrm{box}} \quad , \quad C_{q}^{(j)} = C_{q}^{(j) \mathrm{box}} \, .
\label{Eq:2HDM_DMDDWilsonCoefficients}
\end{align}
We notice that the magnitude of box-diagram contributions depends essentially on the Yukawa-like coupling of the two pseudoscalar mediators with the DM and the SM quarks. On the contrary, the contribution from triangle-shaped diagrams depends also on the trilinear couplings between the CP-even scalars with two pseudoscalar states. The expressions for the Wilson coefficients are rather lengthy and complex, consequently we have summarized them in Appendix~\ref{App:DirectdetectionWC} together with their relation with the \eDMEFT parameters.

From the effective Lagrangian in Eq.~\eqref{Eq:DMDDeffectiveLagrangianWithTwist2Operator} one finds the dimensionless parameter
\begin{align}
C_{N} = m_{N} \left[ \sum_{q=u, d, s} C_{q} f_{T_{q}}^{N} + C_{G} f_{T_{G}}^{N} + \frac{3}{4} \sum_{q\neq t} \left( m_{\chi} C_{Q}^{(1)} + m_{\chi}^{2} C_{Q}^{(2)} \right) \left( q^{N}\left( 2 \right) + \overline{q}^{N} \left( 2 \right) \right) \right]
\end{align}
for the SI~cross section for DM scattering off nucleons given in Eq.~(\ref{Eq:DMDDSIcrossSection}). The nucleon form factors~$f_{q}^{N},f_{TG}^{N}$ have been defined above, whereas~$q^N(2),\overline{q}^N(2)$ are the form factors associated to the twist-2 operators, specified by 
\begin{align}
\langle N \vert \mathcal{O}_{\mu \nu}^{q} \vert N \rangle &= \frac{1}{m_{N}} \left( p_{\mu} p_{\nu} - \frac{1}{4}m_{N}^{2} g_{\mu \nu} \right) \left( q^{N}\left( 2 \right) + \overline{q}^{N} \left( 2 \right) \right) \, .
\end{align}

\begin{figure}[b!]
    \centering
    \includegraphics[width=0.49\textwidth]{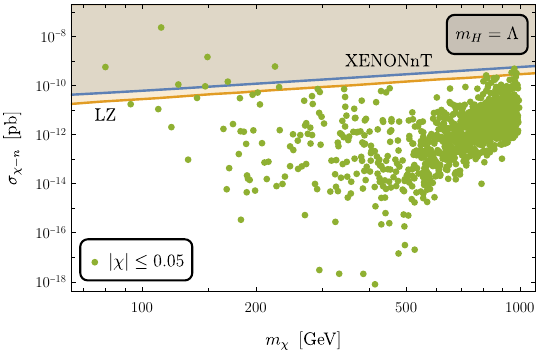}
    \includegraphics[width=0.49\textwidth]{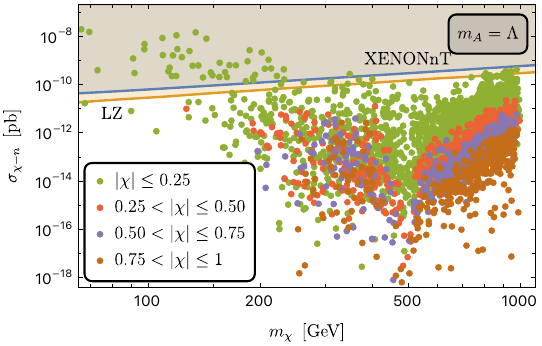}
    \caption{Comparison of the spin-independent DM-nucleon cross sections in the \eDMEFT and the two scenarios in the UV model. The relative difference~$\chi \equiv \left( \sigma_{\chi-n}^{\mathrm{EFT}} - \sigma_{\chi-n}^{\mathrm{UV}} \right)/\sigma_{\chi-n}^{\mathrm{UV}}$ is color-encoded, while the cross sections are confronted with current XENONnT~\cite{XENON:2023cxc} and LZ~\cite{LZ:2022lsv} constraints.}
    \label{Fig:2HDMa_DMDD_results}
\end{figure}

In a similar fashion as for the DM relic density, we compare the predictions for the SI DMDD cross section of the 2HDM+a and different variants of the EFT framework in Fig.~\ref{Fig:2HDMa_DMDD_results}. The parameter points, gathered from a parameter scan over the ranges given above, are displayed in the $(m_\chi,\sigma_{\chi-n})$~plane. The different colors of the point correspond to different values of the relative difference~$\chi \equiv \left( \sigma_{\chi-n}^{\mathrm{EFT}} - \sigma_{\chi-n}^{\mathrm{UV}} \right)/\sigma_{\chi-n}^{\mathrm{UV}}$. The left panel of Fig.~\ref{Fig:2HDMa_DMDD_results} shows again that the EFT realization having two pseudoscalars mediators as propagating degrees of freedom is successful in reproducing the phenomenology of the UV model as all the model points have values of $\chi$ close to zero.

\begin{figure}[b!]
    \centering
    \includegraphics[width=0.55\textwidth]{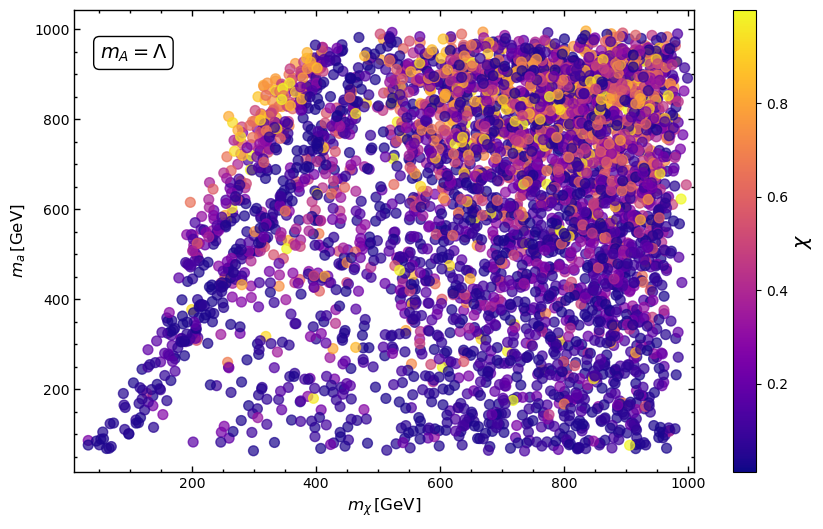}    \caption{Comparison of the spin-independent DM-nucleon cross section in the UV model and the \eDMEFT after integrating out the pseudoscalar $A$, showing the relative differences as a function of DM mass and the mass of the non-integrated pseudoscalar.}
    \label{Fig:2HDMa_DMDD_ma_mChi}
\end{figure}

While the relative difference between the cross sections in the UV model and \eDMEFT are negligibly small in the scenario in which the two pseudoscalars~$A,a$ are propagating degrees of freedom, they can become sizable in regions of parameter space in the scenario with $m_{A}=\Lambda$.  The reason is based on the fact that, in most of the parameter space, the DMDD cross section is dominated by the contribution from the Feynman diagrams with triangle topologies. The triangle diagrams in which the $H$-state is exchanged in the t-channel in general give a small contribution to the DM scattering cross section as it is suppressed by the mass $M_H$. Integrating out the CP-even state $H$ has hence a negligible impact to the DM scattering cross section in the 2HDM+a which, consequently, can be always successfully compared with the one of a \eDMEFT with two pseudoscalar mediators. On the other hand, making the pseudoscalar states $a,A$, which run inside the triangle loops, heavy leads to different conclusions. 
While for small $m_a$ (and $m_A = \Lambda$) we observe in general a good agreement, since the process is dominated by the light pseudoscalar in the loop, when the gap between the pseudoscalar masses becomes small, i.e. for heavy $m_a\to \Lambda$, the validity of the EFT becomes worse, reaching $|\chi\gtrsim 50\%|$, as can be seen in Fig.~\ref{Fig:2HDMa_DMDD_ma_mChi}.  This trend is again to be expected because states not present in the EFT contribute with a bigger relative importance in this limit, as the overall cross section becomes smaller.
So the \eDMEFT just provides a good description for not too heavy~$a$.

\subsubsection{Collider signatures}

\begin{figure}[b!]
    \centering
    \includegraphics[width=0.49\textwidth]{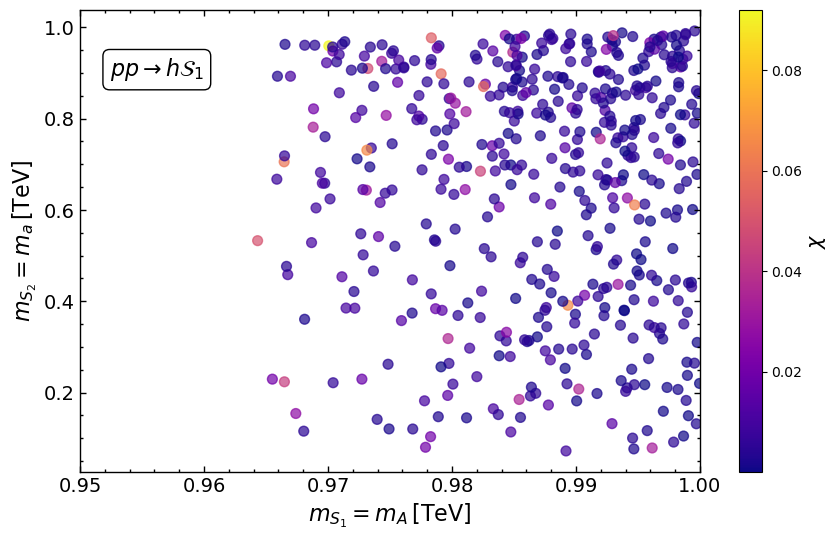}
    \includegraphics[width=0.49\textwidth]{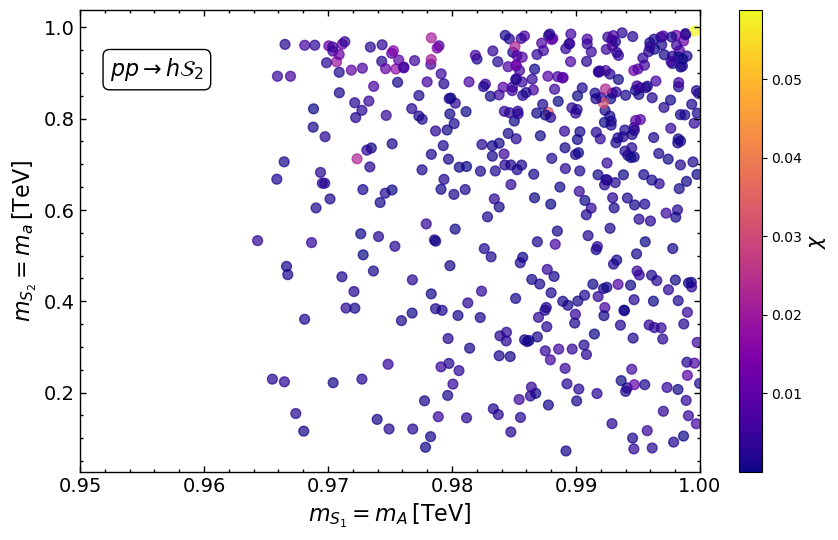}\\
    \includegraphics[width=0.49\textwidth]{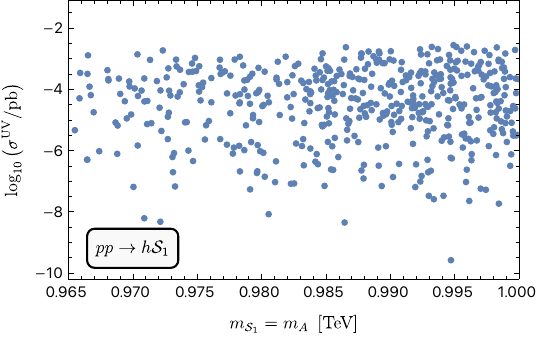}
    \includegraphics[width=0.49\textwidth]{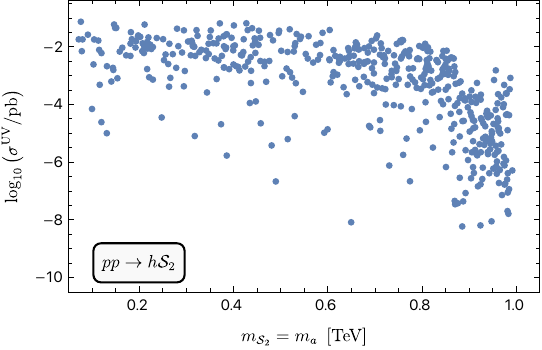}
    \caption{Comparison of the cross sections between the UV models and the EFT for mono-$h$~(\emph{upper row}) and the absolute cross sections~(\emph{lower row}) at LHC with $\sqrt{s}=13\TeV$ center-of-mass energy and $m_{H} = m_{H^{\pm}} = 1\TeV$. The relative deviation is defined as $\chi \equiv \left( \sigma_{pp}^{\mathrm{EFT}} - \sigma_{pp}^{\mathrm{UV}} \right)/\sigma_{pp}^{\mathrm{UV}}$.}
    \label{Fig:2HDMa_crossx}
\end{figure}

Again, we conclude our comparison by analyzing collider phenomenology, focusing on the mono-$h$ signature and the case of $m_H = \Lambda$ (see below). Following a similar strategy as for DM relic density and DMDD, we have performed a parameter scan to compute the production cross section as function of the relevant model parameters. For the 2HDM+a, a very promising maximal value of
\begin{align}
    \sigma_{\mathrm{max}} =
        7.2 \times 10^{-2}\pb 
\end{align}
emerged. As anticipated, the mono-$h$ cross section is significantly larger, compared to the complex-singlet model, due to the discussed resonant enhancement.
The results for the comparison between the UV model and the \eDMEFT are shown in the upper panel of Fig.~\ref{Fig:2HDMa_crossx} in bidimensional planes of the masses of the BSM states involved, while the lower panel contains the absolute cross sections.
As before, the parameter points have been marked with different colors corresponding to the relative difference between the cross sections in the 2HDM+a and the \eDMEFT, finding in general a very good agreement. It is important to stress here, that this is also due to the fact that the scalar $H$, missing in the spectrum of the \eDMEFT, does not contribute to the process in a relevant way in the full 2HDM+a -- given the Higgs does not couple to a scalar-pseudoscalar combination. So even though the validity of the EFT in general is questionable due to the small splitting between the masses of the active $A$ and the integrated $H$ in this last example -- the Lagrangian (\ref{Eq:eDMEFT_Lagrangian}) still reproduces well the full theory in the process at hand (and the exchanged momenta do not actually need to be cut significantly below $\Lambda$). 

The expected number of events for a given luminosity can again be estimated with the formula in Eq.~(\ref{Eq:NumberOfEventsAtColliders}). The above number leads to $\mathcal{O}(10^{4})\times \mathrm{BR}(s \rightarrow \chi \overline{\chi})$ mono-$h$ events at the LHC and one order of magnitude more at the end of the HL-LHC campaign.

We close this section, noting that for the considered hierarchy of $\Lambda=m_H \gg m_A$ the mono-$Z$ signature is always below the experimental sensitivity for center-of-mass energies below $\Lambda$ and cannot be sizable in the \eDMEFT (assuming CP conservation). This is because it requires a scalar and a pseudoscalar degree of freedom interacting with the $Z$ boson.
For a sizable cross section, thus the inverted hierarchy of $m_H\ll m_A=\Lambda$ would need to be considered, leading in turn to a less promising mono-$h$ process. To allow for both processes being sizable at a time, a further-extended 3-mediator \eDMEFT would be required.

\section{Conclusions}
\label{sec:concl}

We have presented a very general framework, dubbed (non-linear) \eDMEFT, for DM phenomenology, based on an  EFT with non-linearly realized electroweak symmetry. As we showed, the approach features various virtues regarding its generality, theoretical consistency, and detectable signatures at collider experiments (exceeding significantly the original simplified models). We have introduced a Lagrangian containing all the gauge-invariant operators up to mass dimension $D=5$, describing the interactions of a fermionic DM candidate and two generic spin-0 mediators with potentially different CP properties. This setups aims at capturing in a simple way the phenomenology of a variety of `UV-complete' dark sector models, including the DM relic density (assuming the standard DM freeze-out mechanism), DM direct detection, and LHC searches for BSM states (with focus on mono-$h$ and mono-$Z$ signatures in this work).
We have verified the effectiveness of the \eDMEFT by comparing its predictions to the corresponding ones from a set of increasingly refined concrete models. Indeed, if a (LHC) signal hinting at a (non-trivial) Dark Sector model is found, the \eDMEFT can be used to parametrize very generally the properties of the mediators and the DM and, with the help of further measurements, to close in on the microscopic nature of the Dark Sector.

\section*{Acknowledgments}
We would like to thank Adam Falkowski, Karla Maria Tame-Narvaez, Jacinto Paulo Neto and Javier Virto for useful discussions. SF would like to thank the Università degli Studi di Messina for their kind hospitality and DCA and GA would like to thank the Max-Planck-Institut f{\"u}r Kernphysik for their kind hospitality. DCA acknowledges funding from the Spanish MCIN/AEI/10.13039/501100011033 through grant PID2022-136224NB-C21.

\clearpage

\begin{appendix}

\section{Wilson coefficients for collider searches and constraints} \label{App:TableWilsonCoefficients}

\begin{table}[b!]
    \centering
    \begin{tabular}{ccc}
        \toprule
        Category  & Process & Model parameter\\
        \midrule
        \multirow{5}*{$\tau$ pair production}  & \begin{tikzpicture}[baseline=7mm]
                    \begin{feynman}
                        \vertex (a1) {\(g\)};
                        \vertex[above=1.5cm of a1] (a2) {\(g\)};

                        \vertex at ($(a1)!0.5!(a2) + (1cm,0)$) (b1) [dot] {};
                        \vertex[right=1cm of b1] (c1);

                        \vertex[right=3cm of a1] (d1) {\(\tau^{+}\)};
                        \vertex[right=3cm of a2] (d2) {\(\tau^{-}\)};

                        \diagram* {
                        (a1) -- [gluon] (b1),
                        (a2) -- [gluon] (b1),
                        (b1) -- [scalar, edge label=\(\mathcal{S}_{i}\)] (c1),
                        (d1) -- [fermion] (c1),
                        (d2) -- [anti fermion] (c1)
                    };
                    \end{feynman}
                \end{tikzpicture}
                 & $c_{G}^{\mathcal{S}_{i}}$,$\tilde{c}_{G}^{\mathcal{S}_{i}}$,$\left( c_{\ell}\right)_{1,0}^{(0)}$,$\left( c_{\ell}\right)_{0,1}^{(0)}$ \\
        & \begin{tikzpicture}[baseline=7mm]
                    \begin{feynman}
                        \vertex (a1) {\(g\)};
                        \vertex[above=1.5cm of a1] (a2) {\(g\)};

                        \vertex[right=1cm of a1] (b1);
                        \vertex[right=1cm of a2] (b2);

                        \vertex at ($(b1)!0.5!(b2) + (0.7cm,0)$) (c1);
                        \vertex[right=1cm of c1] (d1);

                        \vertex at ($(a1) + (3.7cm,0)$) (e1) {\(\tau^{+}\)};
                        \vertex at ($(a2) + (3.7cm,0)$) (e2) {\(\tau^{-}\)};

                        \diagram* {
                        (a1) -- [gluon] (b1),
                        (a2) -- [gluon] (b2),
                        (b1) -- [fermion,edge label=\(q\)] (b2),
                        (b2) -- [fermion] (c1),
                        (c1) -- [fermion] (b1),
                        (c1) -- [scalar, edge label=\(\mathcal{S}_{i}\)] (d1),
                        (e1) -- [fermion] (d1),
                        (e2) -- [anti fermion] (d1)
                    };
                    \end{feynman}
                \end{tikzpicture}
 & $+ \left( c_{q}\right)_{1,0}^{(0)}$,$\left( c_{q}\right)_{0,1}^{(0)}$ \\
        \midrule
        \multirow{3}*{Mono-jet}  & \begin{tikzpicture}[baseline=9mm]
                    \begin{feynman}
                        \vertex (a1) {\(g\)};
                        \vertex[above=1.5cm of a1] (a2) {\(g\)};

                        \vertex at ($(a1)!0.5!(a2) + (1cm,0)$) (b1) [dot] {};

                        \vertex at ($(a1)!0.6!(b1)$) (j1);
                        \vertex at ($(j1) + (1.3cm,-0.3cm)$) (j2) {\(g\)};
                        
                        \vertex[right=1cm of b1] (c1);

                        \vertex[right=3cm of a1] (d1) {\(\overline{\chi}\)};
                        \vertex[right=3cm of a2] (d2) {\(\chi\)};

                        \diagram* {
                        (a1) -- [gluon] (b1),
                        (j2) -- [gluon] (j1),
                        (a2) -- [gluon] (b1),
                        (b1) -- [scalar, edge label=\(\mathcal{S}_{i}\)] (c1),
                        (d1) -- [fermion] (c1),
                        (d2) -- [anti fermion] (c1)
                    };
                    \end{feynman}
                \end{tikzpicture}
                & \multirow{3}*{$c_{G}^{\mathcal{S}_{i}}$,$\tilde{c}_{G}^{\mathcal{S}_{i}}$, $y_{1,0}^{(0)}$, $y_{0,1}^{(0)} + q\text{-}\mathrm{loop}$} \\
          & $+$ $q$-loop & \\
        \midrule
        \multirow{3}*{Higgs pair production}  & \begin{tikzpicture}[baseline=9mm]
                    \begin{feynman}
                        \vertex (a1) {\(g\)};
                        \vertex[above=1.5cm of a1] (a2) {\(g\)};

                        \vertex at ($(a1)!0.5!(a2) + (1cm,0)$) (b1) [dot] {};
                        \vertex[right=1cm of b1] (c1);

                        \vertex[right=3cm of a1] (d1) {\(h\)};
                        \vertex[right=3cm of a2] (d2) {\(h\)};

                        \diagram* {
                        (a1) -- [gluon] (b1),
                        (a2) -- [gluon] (b1),
                        (b1) -- [scalar, edge label=\(\mathcal{S}_{i}\)] (c1),
                        (d1) -- [scalar] (c1),
                        (d2) -- [scalar] (c1)
                    };
                    \end{feynman}
                \end{tikzpicture} & \multirow{3}*{$c_{G}^{\mathcal{S}_{i}}$,$\tilde{c}_{G}^{\mathcal{S}_{i}}$, $\lambda_{1,0}^{(2)}$, $\lambda_{0,1}^{(2)} + q\text{-}\mathrm{loop}$}\\ & $+$ $q$-loop &  \\
        \midrule
        \multirow{3}*{Mono-$h$}  & 
                \begin{tikzpicture}[baseline=9mm]
                    \begin{feynman}
                        \vertex (a1) {\(g\)};
                        \vertex[above=1.5cm of a1] (a2) {\(g\)};

                        \vertex at ($(a1)!0.5!(a2) + (1cm,0)$) (b1) [dot] {};
                        \vertex[right=1cm of b1] (c1);

                        \vertex[right=3cm of a1] (d1) {\(h\)};
                        \vertex[right=3cm of a2] (d2) {\(\mathcal{S}_{i}\)};

                        \diagram* {
                        (a1) -- [gluon] (b1),
                        (a2) -- [gluon] (b1),
                        (b1) -- [scalar, edge label=\(h/\mathcal{S}_{i}\)] (c1),
                        (d1) -- [scalar] (c1),
                        (d2) -- [scalar] (c1)
                    };
                    \end{feynman}
                \end{tikzpicture} & \multirow{3}*{\makecell{$c_{G}^{\mathcal{S}_{i}}$,$\tilde{c}_{G}^{\mathcal{S}_{i}}$,$\lambda_{1,0}^{(2)}$, $\lambda_{0,1}^{(2)}$,$\lambda_{2,0}^{(1)}$,$\lambda_{1,1}^{(1)}$,$\lambda_{0,2}^{(1)}$,\\
                 $+ q\text{-}\mathrm{loop}$}}\\
                
                 & $+$ $q$-loop & \\
                \midrule
                \multirow{3}*{Di-boson resonance}& \begin{tikzpicture}[baseline=9mm]
                    \begin{feynman}
                        \vertex (a1) {\(g\)};
                        \vertex[above=1.5cm of a1] (a2) {\(g\)};

                        \vertex at ($(a1)!0.5!(a2) + (1cm,0)$) (b1) [dot] {};
                        \vertex[right=1cm of b1] (c1);

                        \vertex[right=3cm of a1] (d1) {\(Z/W^{-}\)};
                        \vertex[right=3cm of a2] (d2) {\(Z/W^{+}\)};

                        \diagram* {
                        (a1) -- [gluon] (b1),
                        (a2) -- [gluon] (b1),
                        (b1) -- [scalar, edge label=\(h/\mathcal{S}_{i}\)] (c1),
                        (d1) -- [boson] (c1),
                        (d2) -- [boson] (c1)
                    };
                    \end{feynman}
                \end{tikzpicture} & \multirow{3}*{\makecell{$c_{G}^{\mathcal{S}_{i}}$,$\tilde{c}_{G}^{\mathcal{S}_{i}}$,$c_{W}^{\mathcal{S}_{i}}$,$c_{B}^{\mathcal{S}_{i}}$,$\kappa_{0,0}^{(1)}$,$\kappa_{1,0}^{(0)}$,$\kappa_{0,1}^{(0)}$\\ $+ q\text{-}\mathrm{loop}$}}\\ & $+$ $q$-loop & \\
                \midrule
                \multirow{3}*{Mono-$Z$}&\begin{tikzpicture}[baseline=9mm]
                    \begin{feynman}
                        \vertex (a1) {\(g\)};
                        \vertex[above=1.5cm of a1] (a2) {\(g\)};

                        \vertex at ($(a1)!0.5!(a2) + (1cm,0)$) (b1) [dot] {};
                        \vertex[right=1cm of b1] (c1);

                        \vertex[right=3cm of a1] (d1) {\(Z\)};
                        \vertex[right=3cm of a2] (d2) {\(h/\mathcal{S}_{i}\)};

                        \diagram* {
                        (a1) -- [gluon] (b1),
                        (a2) -- [gluon] (b1),
                        (b1) -- [scalar, edge label=\(h/\mathcal{S}_{i}\)] (c1),
                        (d1) -- [boson] (c1),
                        (d2) -- [scalar] (c1)
                    };
                    \end{feynman}
                \end{tikzpicture}& \multirow{2}*{\makecell{$c_{G}^{\mathcal{S}_{i}}$,$\tilde{c}_{G}^{\mathcal{S}_{i}}$,$h_{1,0}^{(0)}$,$h_{0,1}^{(0)}$,$s1_{0,0}^{(1)}$,\\
                $s1_{0,1}^{(0)}$,$s2_{0,0}^{(1)}$,$s2_{1,0}^{(0)}$\\
                $+ q\text{-}\mathrm{loop}$}}
                \\
                & $+$ $q$-loop & \\
                \bottomrule
    \end{tabular}
    \caption{Wilson coefficients for collider searches. The tilde indicates that the coupling involves a pseudoscalar mediator.
    }
    \label{Tab:ListWilsonCoefficients_ColliderSearches}
\end{table}

\clearpage

\begin{table}[h!]
    \centering
    \begin{tabular}{ccc}
        \toprule
        Category  & Process & Model parameter \\
        \midrule
        \multirow{6}*{DMDD}  & 
                \begin{tikzpicture}[baseline=9mm]
                    \begin{feynman}
                        \vertex (a1) {\(q\)};
                        \vertex[right=2cm of a1] (a2) {\(q\)};
                        \vertex[above=2cm of a1] (e1) {\(\chi\)};
                        \vertex[above=2cm of a2] (e2) {\(\chi\)};

                        \vertex at ($(a1)!0.5!(a2) + (a1)!0.2!(e1)$) (b1);
                        \vertex at ($(a1)!0.5!(a2) + (e1)!0.2!(a1)$) (d1);

                        \diagram* {
                        (a1) -- [fermion] (b1),
                        (a2) -- [anti fermion] (b1),
                        (b1) -- [scalar] (d1),  
                        (d1) -- [anti fermion] (e1),
                        (d1) -- [fermion] (e2)
                    };
                    \end{feynman}
                \end{tikzpicture} & $y_{0,0}^{(1)}$, $y_{1,0}^{(0)}$, $y_{0,1}^{(0)}$, $\left( c_{q}\right)_{0,0}^{(1)}$, $\left( c_{q}\right)_{1,0}^{(0)}$, $\left( c_{q}\right)_{0,1}^{(0)}$ \\
          &
                \begin{tikzpicture}[baseline=9mm]
                    \begin{feynman}
                        \vertex (a1) {\(g\)};
                        \vertex[right=2cm of a1] (a2) {\(g\)};
                        \vertex[above=2cm of a1] (e1) {\(\chi\)};
                        \vertex[above=2cm of a2] (e2) {\(\chi\)};

                        \vertex at ($(a1)!0.5!(a2) + (a1)!0.2!(e1)$)[dot] (b1) {};
                        \vertex at ($(a1)!0.5!(a2) + (e1)!0.2!(a1)$) (d1);

                        \diagram* {
                        (a1) -- [gluon] (b1),
                        (a2) -- [gluon] (b1),
                        (b1) -- [scalar] (d1),  
                        (d1) -- [anti fermion] (e1),
                        (d1) -- [fermion] (e2)
                    };
                    \end{feynman}
                    \end{tikzpicture} & $c_{G}^{h}$, $c_{G}^{\mathcal{S}_{i}}$ \\
        \midrule
        \multirow{6}*{DM}  & 
                \begin{tikzpicture}[baseline=9mm]
                    \begin{feynman}
                        \vertex (a1) {\(\overline{\chi}\)};
                        \vertex[above=1.5cm of a1] (a2) {\(\chi\)};

                        \vertex at ($(a1)!0.5!(a2) + (1cm,0)$) (b1);
                        \vertex[right=1cm of b1] (c1) {\(h/\mathcal{S}_{i}\)};

                        \diagram* {
                        (a1) -- [anti fermion] (b1),
                        (a2) -- [fermion] (b1),
                        (b1) -- [scalar] (c1)
                    };
                    \end{feynman}
                \end{tikzpicture} & $y_{0,0}^{(1)}$, $y_{1,0}^{(0)}$, $y_{0,1}^{(0)}$ \\
          & 
                \begin{tikzpicture}[baseline=9mm]
                    \begin{feynman}
                        \vertex (a1) {\(\overline{\chi}\)};
                        \vertex[above=1.5cm of a1] (a2) {\(\chi\)};

                        \vertex at ($(a1)!0.5!(a2) + (1cm,0)$) (b1) [dot] {};
                        
                        \vertex[right=2cm of a1] (c1) {\(h/\mathcal{S}_{i}\)};
                        \vertex[above=1.5cm of c1] (c2) {\(h/\mathcal{S}_{i}\)};

                        \diagram* {
                        (a1) -- [anti fermion] (b1),
                        (a2) -- [fermion] (b1),
                        (b1) -- [scalar] (c1),
                        (b1) -- [scalar] (c2)
                    };
                    \end{feynman}
                \end{tikzpicture} & $y_{0,0}^{(2)}$, $y_{2,0}^{(0)}$, $y_{0,2}^{(0)},y_{1,0}^{(1)}$, $y_{0,1}^{(1)}$, $y_{1,1}^{(0)}$ \\
        \midrule
        IHD  &  \begin{tikzpicture}[baseline=9mm]
                    \begin{feynman}
                        \vertex (a1) {\(\mathcal{S}_{i}\)};
                        \vertex[above=2cm of a1] (a2) {\(\mathcal{S}_{i}\)};
                        \filldraw[black] (0.3,1.5) circle (1pt) {};
                        \filldraw[black] (0.4,1) circle (1pt) {};
                        \filldraw[black] (0.3,0.5) circle (1pt) {};

                        \vertex at ($(a1)!0.5!(a2) - (1cm,0)$) (b1);
                        
                        \vertex[left=1cm of b1] (c1) {\(h\)};

                        \diagram* {
                        (a1) -- [scalar] (b1),
                        (a2) -- [scalar] (b1),
                        (b1) -- [scalar] (c1)
                    };
                    \end{feynman}
                \end{tikzpicture} & \makecell{$\lambda_{2,0}^{(1)}$, $\lambda_{1,1}^{(1)}$, $\lambda_{0,2}^{(1)}$, $\lambda_{3,0}^{(1)}$, $\lambda_{2,1}^{(1)}$, $\lambda_{1,2}^{(1)}$, $\lambda_{0,3}^{(1)}$, \\ \\ $\lambda_{4,0}^{(1)}$, $\lambda_{3,1}^{(1)}$, $\lambda_{2,2}^{(1)}$, $\lambda_{1,3}^{(1)}$, $\lambda_{0,4}^{(1)}$} \\
        \midrule
        \multirow{9}*{EWPO}  & 
                \begin{tikzpicture}[baseline=6mm]
                    \begin{feynman}
                        \vertex (a1) {\(\mathcal{S}_{i}\)};
                        \vertex[above=1.5cm of a1] (a2) {\(\mathcal{S}_{i}\)};
                        \filldraw[black] (0.3,1.1) circle (1pt) {};
                        \filldraw[black] (0.4,0.75) circle (1pt) {};
                        \filldraw[black] (0.3,0.4) circle (1pt) {};

                        \vertex at ($(a1)!0.5!(a2) - (1cm,0)$) (b1);
                        
                        \vertex[left=1cm of b1] (c1) {\(Z\)};

                        \diagram* {
                        (a1) -- [scalar] (b1),
                        (a2) -- [scalar] (b1),
                        (b1) -- [boson] (c1)
                    };
                    \end{feynman}
                \end{tikzpicture} & $s1_{1,0}^{(0)}$, $s1_{0,1}^{(0)}$, $s1_{2,0}^{(0)}$, $s1_{1,1}^{(0)}$, $s1_{0,2}^{(0)}$ + $\left( s1 \leftrightarrow s2 \right)$ \\
          & 
                \begin{tikzpicture}[baseline=3mm]
                    \begin{feynman}
                        \vertex (a1) {\(V\)};
                        \vertex[right=1cm of a1] (a2);
                        \vertex[right=0.00001cm of a2] (xxx); 
                        \vertex[right=1cm of a2] (a3) {\(V\)};

                        \diagram* {
                        (a1) -- [boson] (a2) -- [scalar, out=135, in=45, min distance=2cm] (xxx) -- [boson] (a3),
                    };
                    \end{feynman}
                \end{tikzpicture} & $\kappa_{0,0}^{(2)}$, $\kappa_{2,0}^{(0)}$, $\kappa_{0,2}^{(0)}$ \\
          & 
                \begin{tikzpicture}[baseline=-1mm]
                    \begin{feynman}
                        \vertex (a1) {\(V\)};
                        \vertex[right=1cm of a1] (a2);
                        \vertex[right=1cm of a2] (a3);
                        \vertex[right=0.8cm of a3] (a4) {\(V\)};

                        \diagram* {
                        (a1) -- [boson] (a2) -- [scalar, out=90, in=90, looseness=1.5] (a3) -- [boson] (a4),
                        (a2) -- [boson, out=-90, in=-90, looseness=1.5] (a3)
                    };
                    \end{feynman}
                \end{tikzpicture} & $\kappa_{0,0}^{(1)}$, $\kappa_{1,0}^{(0)}$, $\kappa_{0,1}^{(0)}$ \\
          & 
                \begin{tikzpicture}[baseline=-1mm]
                    \begin{feynman}
                        \vertex (a1) {\(V\)};
                        \vertex[right=1cm of a1] (a2);
                        \vertex[right=1cm of a2] (a3);
                        \vertex[right=0.8cm of a3] (a4) {\(V\)};

                        \diagram* {
                        (a1) -- [boson] (a2) -- [scalar, out=90, in=90, looseness=1.5] (a3) -- [boson] (a4),
                        (a2) -- [scalar, out=-90, in=-90, looseness=1.5] (a3)
                    };
                    \end{feynman}
                \end{tikzpicture} & \makecell{$h_{0,0}^{(1)}$, $h_{1,0}^{(0)}$, $h_{0,1}^{(0)}$ + $\left( s \leftrightarrow s1 \leftrightarrow s2 \right)$ } \\
                \bottomrule
    \end{tabular}
    \caption{Important model parameters for various categories of constraints: dark matter direct detection (DMDD), DM annihilation, invisible SM Higgs decay (IHD), and EW precision observables (EWPO). A dot in the vertex indicates an effective vertex. If not indicated otherwise, dashed propagators correspond to the SM Higgs~$h$ and BSM scalars~$\mathcal{S}_{i}$.}
    \label{Tab:ListWilsonCoefficients_Constraints}
\end{table}

\clearpage

\section{Theoretical and experimental constraints for UV models}

\subsection{SM + one complex scalar singlet} \label{App:OneSingletExtension_Constraints}
\subsubsection*{Vacuum stability}
The potential can be written as
\begin{align}
    -\mathcal{L} \supset \frac{\lambda_{\Phi}}{4} \left( h^{2} + \frac{\lambda_{\Phi S}}{2\lambda_{\Phi}} s^{2} \right)^{2} + \frac{4\lambda_{\Phi}^{2}\lambda_{S} - \lambda_{\Phi S}^{2} \lambda_{\Phi}}{16\lambda_{\Phi}^{2}} s^{4} + \lambda_{\Phi}v h^{3} + \lambda_{s}v s^{3}
\end{align}
and the constraints for the potential being bounded from below read
\begin{align}
    \lambda_{\Phi} > 0 \, , \, \lambda_{s} > 0 \, , \, 4\lambda_{\Phi} \lambda_{S} > \lambda_{\Phi S}^{2} \, .
\end{align}
\subsubsection*{Perturbativity and perturbative unitarity}
\begin{align}
    Q=2: \quad &\left( G^{+} G^{+}/\sqrt{2} \right) \\
    Q=1: \quad &\left( G^{+} h \right),\left( G^{+} G^{0} \right),\left( G^{+} s \right),\left( G^{+} a \right) \\
    Q=0: \quad & \left( G^{+} G^{-} \right), \left( G^{0}G^{0}/\sqrt{2} \right), \left( hh/\sqrt{2} \right), \left( ss/\sqrt{2} \right), \left( aa/\sqrt{2} \right), \nonumber \\ 
    &\left( h G^{0} \right), \left( hs \right), \left( ha \right), \left( G^{0}s \right), \left( G^{0}a \right),\left( sa \right)
\end{align}
The Hermitian scattering matrices~$\mathcal{M}_{Q}$ with the electric charge~$Q$ read
\begin{align}
    \mathcal{M}_{2} &= \mathrm{diag} \left( 2\lambda_{\Phi} \right)\\
    \mathcal{M}_{1} &= \mathrm{diag} \left( 2\lambda_{\Phi}, 2\lambda_{\Phi}, \lambda_{\Phi S}, \lambda_{\Phi S} \right)\\
    \mathcal{M}_{0} &= \mathrm{diag} \left( A_{5\times 5}, 2\lambda_{\Phi}, \lambda_{\Phi S}, \lambda_{\Phi S}, \lambda_{\Phi S}, \lambda_{\Phi S}, 2\lambda_{\Phi S} \right)
\end{align}
with the submatrix
\begin{align}
    A_{5\times 5} = \begin{pmatrix}
        4\lambda_{\Phi} & \times & \times & \times & \times \\
        \sqrt{2} \lambda_{\Phi} & 3\lambda_{\Phi} & \times & \times & \times \\
        \sqrt{2} \lambda_{\Phi} & \lambda_{\Phi} & 3\lambda_{\Phi} & \times & \times \\
        \lambda_{\Phi S}/\sqrt{2} & \lambda_{\Phi S}/2 & \lambda_{\Phi S}/2 & 3\lambda_{S} & \times \\
        \lambda_{\Phi S}/\sqrt{2} & \lambda_{\Phi S}/2 & \lambda_{\Phi S}/2 & \lambda_{S} & 3\lambda_{S}
    \end{pmatrix} \, .
\end{align}
\subsubsection*{Invisible Higgs decay}
\begin{align}
    \Gamma \left( h \rightarrow ss \right) &= \frac{\left\vert g_{hss} \right\vert^{2}}{32\pi m_{h}} \sqrt{1 - \frac{4m_{s}^{2}}{m_{h}^{2}}} \\
    \Gamma \left( h \rightarrow aa \right) &= \frac{\left\vert g_{haa} \right\vert^{2}}{32\pi m_{h}} \sqrt{1 - \frac{4m_{a}^{2}}{m_{h}^{2}}}
\end{align}
with the couplings
\begin{align}
    g_{hss} &= \lambda_{\Phi S} \left( c_{\theta}^{3} v_{h} + s_{\theta}^{3} v_{s} \right) - s_{2\theta} \lambda_{\Phi S} \left( v_{s} c_{\theta} + v_{h} s_{\theta} \right) + 3 s_{2\theta} \left( \lambda_{S} v_{s} c_{\theta} + \lambda_{\Phi} v_{h} s_{\theta} \right) \\    
    g_{haa} &= 2\lambda_{S} s_{\theta} v_{s} + c_{\theta} \lambda_{\Phi S} v_{h} \, .
\end{align}

\subsection{SM + two scalar singlets} \label{App:TwoSingletExtension_Constraints}
\subsubsection*{Vacuum stability}
The potential can be cast as
\begin{align}
    -\mathcal{L} &\supset \frac{\lambda_{\Phi}}{8} \left[ \left( h^{2} + \frac{\lambda_{\Phi 1}}{\lambda_{\Phi}} \rho_{1}^{2} \right)^{2} + \left( h^{2} + 2 \frac{\lambda_{\Phi 2}}{\lambda_{\Phi}} \rho_{2}^{2} \right)^{2} \right] + \lambda_{\Phi} v h^{3} + \lambda_{1} v_{1} \rho_{1}^{3} + 4\lambda_{2} v_{2} \rho_{2}^{3} \nonumber \\
    &\hspace{-5mm} + \frac{2\lambda_{1}\lambda_{\Phi} - \lambda_{\Phi 1}^{2}}{8 \lambda_{\Phi}} \left( \rho_{1}^{2} + \frac{2\lambda_{\Phi} \lambda_{3}}{2\lambda_{1}\lambda_{\Phi} - \lambda_{\Phi 1}^{2}} \rho_{2}^{2} \right)^{2} + \frac{\left( 2\lambda_{1}\lambda_{\Phi} - \lambda_{\Phi 1}^{2} \right)\left( 2\lambda_{2}\lambda_{\Phi} - \lambda_{\Phi 2}^{2} \right) - \lambda_{\Phi}^{2} \lambda_{3}^{2}}{2\lambda_{\Phi} \left( 2\lambda_{1}\lambda_{\Phi} - \lambda_{\Phi 1}^{2} \right)} \rho_{2}^{4} \, ,
\end{align}
such that the vacuum stability in the directions of the three scalars is given by
\begin{align}
    \lambda_{\Phi,1} >0 \quad , \quad 2\lambda_{1}\lambda_{\Phi} > \lambda_{\Phi 1}^{2} \quad , \quad \left( 2\lambda_{1}\lambda_{\Phi} - \lambda_{\Phi 1}^{2} \right) \left( 2\lambda_{2}\lambda_{\Phi} - \lambda_{\Phi 2}^{2} \right) > \lambda_{\Phi}^{2} \lambda_{3}^{2} \, .
\end{align}
\subsubsection*{Perturbativity and perturbative unitarity}
Perturbative unitarity restricts the couplings in two-to-two scalar interactions. The general expression of the scattering amplitude
\begin{align}
    \mathcal{M}^{(kf)} = 16\pi i \sum_{J \geq 0} \left( 2J +1 \right) a_{J}^{(kf)} \left( s \right) P_{J} \left( \cos \theta \right)
\end{align}
with the indices~$k,f$ for the initial and final states, respectively, the Legendre polynomials~$P_{J}$ and the scattering angle~$\theta$ simplifies in the high-energy limit, such that
\begin{align}
    a_{0}^{(kf)} = -\frac{i}{16\pi} \mathcal{M}^{(kf)} \quad \mathrm{and} \quad a_{J}^{(kf)} = 0 \quad \forall J \geq 1 \, . 
\end{align}
Following the discussion in~\cite{Arhrib:2011uy}, we demand $\vert \mathrm{Re} \, a_{0}^{(kf)} \vert \leq 1/2$ by restricting the eigenvalues.
\begin{align}
    Q=2: \quad &\left( G^{+} G^{+}/\sqrt{2} \right) \\
    Q=1: \quad &\left( G^{+} h \right),\left( G^{+} G^{0} \right),\left( G^{+} s_{1} \right),\left( G^{+} a_{1} \right),\left( G^{+} s_{2} \right) \\
    Q=0: \quad & \left( G^{+} G^{-} \right), \left( G^{0}G^{0}/\sqrt{2} \right), \left( hh/\sqrt{2} \right), \left( s_{1}s_{1}/\sqrt{2} \right), \left( a_{1}a_{1}/\sqrt{2} \right), \nonumber \\ 
    &\left( s_{2}s_{2}/\sqrt{2} \right), \left( h G^{0} \right), \left( hs_{1} \right), \left( ha_{1} \right), \left( G^{0}s_{1} \right), \left( G^{0}a_{1} \right), \left( hs_{2} \right), \nonumber \\ 
    &\left( G^{0}s_{2} \right), \left( s_{1}a_{1} \right), \left( s_{1}s_{2} \right), \left( a_{1}s_{2} \right)
\end{align}
The Hermitian scattering matrices~$\mathcal{M}_{Q}$ with the electric charge~$Q$ read
\begin{align}
    \mathcal{M}_{2} &= \mathrm{diag} \left( 2\lambda_{\Phi} \right)\\
    \mathcal{M}_{1} &= \mathrm{diag} \left( 2\lambda_{\Phi}, 2\lambda_{\Phi}, \lambda_{\Phi 1}, \lambda_{\Phi 1}, 2\lambda_{\Phi 2} \right)\\
    \mathcal{M}_{0} &= \mathrm{diag} \left( A_{6\times 6}, 2\lambda_{\Phi}, \lambda_{\Phi 1}, \lambda_{\Phi 1}, \lambda_{\Phi 1}, \lambda_{\Phi 1}, 2\lambda_{\Phi 2}, 2\lambda_{\Phi 2}, 2\lambda_{1}, 2\lambda_{3}, 2\lambda_{3} \right)
\end{align}
with the submatrix
\begin{align}
    A_{6\times 6} = \begin{pmatrix}
        4\lambda_{\Phi} & \times & \times & \times & \times & \times \\
        \sqrt{2} \lambda_{\Phi} & 3\lambda_{\Phi} & \times  & \times & \times & \times \\
        \sqrt{2} \lambda_{\Phi} & \lambda_{\Phi} & 3\lambda_{\Phi} & \times & \times & \times \\
        \lambda_{\Phi 1}/\sqrt{2} & \lambda_{\Phi 1}/2 & \lambda_{\Phi 1}/2 & 3\lambda_{1} & \times & \times \\
        \lambda_{\Phi 1}/\sqrt{2} & \lambda_{\Phi 1}/2 & \lambda_{\Phi 1}/2 & \lambda_{1} & 3\lambda_{1} & \times \\
        \sqrt{2}\lambda_{\Phi 2} & \lambda_{\Phi 2} & \lambda_{\Phi 2} & \lambda_{3} & \lambda_{3} & 12\lambda_{2}
    \end{pmatrix} \, .
\end{align}

\subsubsection*{Invisible Higgs decay}
\begin{align}
    \Gamma \left( h \rightarrow s_{1}s_{1} \right) &= \frac{\left\vert g_{hs_{1}s_{1}} \right\vert^{2}}{32\pi m_{h}} \sqrt{1 - \frac{4m_{s_{1}}^{2}}{m_{h}^{2}}} \\
    \Gamma \left( h \rightarrow s_{1}s_{2} \right) &= \frac{\left\vert g_{hs_{1}s_{2}} \right\vert^{2}}{16\pi m_{h}} \sqrt{\left( m_{h}^{2} - \left( m_{s_{1}} - m_{s_{2}} \right)^{2} \right)\left( m_{h}^{2} - \left( m_{s_{1}} + m_{s_{2}} \right)^{2} \right)} \\
    \Gamma \left( h \rightarrow s_{2}s_{2} \right) &= \frac{\left\vert g_{hs_{2}s_{2}} \right\vert^{2}}{32\pi m_{h}} \sqrt{1 - \frac{4 m_{s_{2}}^{2}}{m_{h}^{2}}} \\
    \Gamma \left( h \rightarrow a_{1}a_{1} \right) &= \frac{\left\vert g_{ha_{1}a_{1}} \right\vert^{2}}{32\pi m_{h}} \sqrt{1 - \frac{4m_{a_{1}}^{2}}{m_{h}^{2}}}
\end{align}
with the couplings
\begin{align}
    g_{hs_{1}s_{1}} &= -\left( c_{\theta}^{2} \lambda_{\Phi 1} + 2\lambda_{\Phi 2} s_{\theta}^{2} \right) v_{h}
    \label{eq:multi_cop_s1s1}\\
    g_{hs_{2}s_{2}} &= -\left( s_{\theta}^{2} \lambda_{\Phi 1} + 2\lambda_{\Phi 2} c_{\theta}^{2} \right) v_{h} \\    
    g_{ha_{1}a_{1}} &= -\lambda_{\Phi 1} v_{h} \\
    g_{hs_{1}s_{2}} &= s_{\theta}c_{\theta} \left( \lambda_{\Phi 1} - 2\lambda_{\Phi 2} \right) v_{h} \, .
    \label{eq:multi_cop_s1s2}
\end{align}

\subsection{2HDM + one singlet} \label{App:2HDMa_Constraints}
The theoretical constraints that we take into account for our numerical analysis ensure perturbative unitarity~\cite{Kanemura:2004mg,Becirevic:2015fmu,Barroso:2013awa,Arcadi:2022lpp} and the potential being bounded from below~\cite{Kannike:2012pe}. 

\subsubsection*{Perturbativity and perturbative unitarity}
Demanding that the eigenvalues of the scattering matrix are bounded from above to preserve perturbative unitarity, the constraints read
\begin{align}
    \left\vert \lambda_{11P} \right\vert \, , \, \left\vert \lambda_{22P} \right\vert \, , \, \left\vert \lambda_{3} \pm \lambda_{4} \right\vert \, , \, \frac{1}{2}\left\vert \lambda_{3} \pm \lambda_{5} \right\vert \, , \, \frac{1}{2}\left\vert \lambda_{3} + 2\lambda_{4} \pm 3\lambda_{5} \right\vert &< 4\pi \\
    \left\vert \frac{1}{2} \left( \lambda_{1} + \lambda_{2} \pm \sqrt{\left( \lambda_{1} - \lambda_{2} \right)^{2} + 4\lambda_{4,5}^{2}} \right) \right\vert \, , \, \left\vert x_{j} \right\vert &< 8\pi
\end{align}
with~$x_{j}$ being the three solutions of the equation
\begin{align}
    0 &= x^{3} - 3\left( \lambda_{P} + \lambda_{1} + \lambda_{2} \right) x^{2} \nonumber \\
    &\hspace{5mm} + \left[ 9\left( \lambda_{1} + \lambda_{2} \right) \lambda_{P} - 4\left( \lambda_{11P}^{2} + \lambda_{22P}^{2} + \lambda_{3}^{2} + \lambda_{3}\lambda_{4} \right) - \lambda_{4}^{2} + 9\lambda_{1}\lambda_{2} \right]x \nonumber \\
    &\hspace{5mm} + 12\left( \lambda_{11P}^{2}\lambda_{2} + \lambda_{22P}^{2}\lambda_{1} \right) - 8\lambda_{11P}\lambda_{22P} \left( 2\lambda_{3} + \lambda_{4} \right) \nonumber \\
    &\hspace{5mm} + 3\lambda_{P}\left( -9\lambda_{1}\lambda_{2} + 4\lambda_{3}^{2} + 4\lambda_{3}\lambda_{4} + \lambda_{4}^{2} \right) \, .
\end{align}

\subsubsection*{Vacuum stability}
Vacuum stability constrains the parameter space by
\begin{align}
    \lambda_{1,2,P} > 0 \quad , \quad \overline{\lambda}_{12} \equiv \lambda_{3} + \sqrt{\lambda_{1}\lambda_{2}} + \min\left( 0, \lambda_{4} - \left\vert \lambda_{5} \right\vert \right) &> 0 \nonumber \\
    \overline{\lambda}_{1P} \equiv \sqrt{\frac{\lambda_{1}\lambda_{P}}{2}} + \lambda_{11P} \quad , \quad \overline{\lambda}_{2P} \equiv \sqrt{\frac{\lambda_{2}\lambda_{P}}{2}} + \lambda_{22P} &> 0 \nonumber \\
    \sqrt{\frac{\lambda_{1} \lambda_{2} \lambda_{P}}{2}} + \lambda_{11P} \sqrt{\lambda_{2}} + \lambda_{22P} \sqrt{\lambda_{1}} + \left( \overline{\lambda}_{12} - \sqrt{\lambda_{1}\lambda_{2}} \right) \sqrt{\frac{\lambda_{P}}{2}} + \sqrt{2\overline{\lambda}_{12}\overline{\lambda}_{1P}\overline{\lambda}_{2P}} &> 0 \, .
\end{align}
As discussed in~\cite{Arcadi:2022lpp}, the coupling parameter~$\lambda_{3}$ is bounded from below for the hierarchy $m_{A} \gg m_{a}$ by
\begin{align}
    \lambda_{3} > \frac{m_{A}^{2} - m_{a}^{2}}{v_h^{2}} \sin^{2} \theta - \frac{m_{h}^{2}}{v_h^{2}} \cot^{2} 2\beta \, ,
\end{align}
which restricts the eigenvalues of the scattering amplitude matrix for this hierarchy to fulfill~\cite{Goncalves:2016iyg}
\begin{align}
    \frac{1}{v_h^{2}} \left\vert \Delta - \frac{m_{A}^{2} - m_{a}^{2}}{8} \left( 1 - \cos 4\theta \right) \pm \sqrt{\Delta^{2} + \frac{\left( m_{A}^{2} - m_{a}^{2} \right)^{2}}{8} \left( 1 - \cos 4\theta \right)} \right\vert < 8\pi
\end{align}
with
\begin{align}
    \Delta \equiv -\frac{2 M_{12}^{2}}{\sin 2\beta} - m_{H^{\pm}}^{2} - \frac{m_{h}^{2}}{2} + 2 m_{W}^{2} \, .
\end{align}

\subsubsection*{Invisible Higgs decay}
The LHC allows to set upper bounds on the branching ratio of the SM Higgs decaying into particles evading detection. The SM predicts~$\Gamma_{h}^{\mathrm{SM}} = 4.1\MeV$ for the decay width and an indirect measurement of the on-shell and off-shell Higgs production cross sections performed by CMS suggests $\Gamma_{h}^{\mathrm{exp}} = 3.1_{-1.7}^{+2.4}\MeV$~\cite{CMS:2022ley}. The branching ratio is $\mathrm{BR} \left( h \rightarrow \mathrm{inv.} \right) < 0.107$~\cite{ATLAS:2023tkt} and the invisible Higgs decay width reads
\begin{align}
    \Gamma_{h}^{\mathrm{BSM}} = \sum_{\phi_{1},\phi_{2}}\Gamma \left( h \rightarrow \phi_{1}\phi_{2} \right)
\end{align}
with kinematically accessible scalars~$\phi_{1,2} \in \{ A,a \}$ and the partial decay widths
\begin{align}
    \Gamma \left( h \rightarrow AA \right) &= \frac{\left\vert g_{hAA} \right\vert^{2}}{32\pi m_{h}} \sqrt{1 - \frac{4 m_{A}^{2}}{m_{h}^{2}}} \\
    \Gamma \left( h \rightarrow aA \right) &= \frac{\left\vert g_{haA} \right\vert^{2}}{16\pi m_{h}^{3}} \sqrt{\left( m_{h}^{2} - \left( m_{a} - m_{A} \right)^{2} \right)\left( m_{h}^{2} - \left( m_{a} + m_{A} \right)^{2} \right)} \\
    \Gamma \left( h \rightarrow aa \right) &= \frac{\left\vert g_{haa} \right\vert^{2}}{32\pi m_{h}} \sqrt{1 - \frac{4 m_{a}^{2}}{m_{h}^{2}}} \, .
\end{align}
The couplings are given by
\begin{align}
    g_{hAA} &= \frac{m_{h}^{2} + 4m_{H^{\pm}}^{2} - 2m_{A}^{2} - 2m_{H}^{2} - 2\lambda_{3} v_h^{2}}{v_h}c_{\theta}^{2} - 2\left( \lambda_{11P} c_{\beta}^{2} + \lambda_{22P} s_{\beta}^{2} \right) s_{\theta}^{2} v_h \\
    g_{haH} &= \frac{s_{2\theta}}{2v_h} \left( m_{h}^{2} + 4m_{H^{\pm}}^{2} - 2m_{H}^{2} - m_{A}^{2} - m_{a}^{2} \right.\nonumber \\
    &\hspace{15mm} \left. + \left( \lambda_{11P} - \lambda_{22P} \right)c_{2\beta} v_h^{2} + \left( \lambda_{11P} + \lambda_{22P} - 2\lambda_{3} \right) v_h^{2} \right)\\
    g_{haa} &= \frac{2m_{a}^{2} + 2m_{H}^{2} - 4m_{H^{\pm}}^{2} - m_{h}^{2} + 2\lambda_{3} v_h^{2}}{v_h} s_{\theta}^{2} + 2\left( \lambda_{11P} c_{\beta}^{2} + \lambda_{22P} s_{\beta}^{2} \right)c_{\theta}^{2} v_h \, .
\end{align}

\subsubsection*{$B$ physics observables}
Measurements of  branching ratios of $B$ mesons put strong constraints on the mass of the electrically charged scalar. The results in 2018 presented by the Heavy Flavor Averaging Group show $\mathrm{BR} \left( b \rightarrow s\gamma \right) = \left( 3.32 \pm 0.15 \right) \cdot 10^{-4}$~\cite{HFLAV:2019otj} which excludes masses~$M_{H^{\pm}} \lesssim 800\GeV$~\cite{Misiak:2020vlo}.\footnote{The authors of this article are aware of the latest measurement~\cite{HFLAV:2022esi} of the branching ratio which suggests a slightly larger value with a larger uncertainty. Consequently, the lower bound on possible~$m_{H^{\pm}}$ is reduced, as discussed in~\cite{Biekotter:2024ykp}. In the present analysis we keep the stronger constraint.}

\subsubsection*{Electroweak precision tests}
The precise measurements of the masses of the massive EW gauge bosons allow to constrain new-physics contributions. The contribution~$\Delta \rho$ to the $\rho$ parameter is given by
\begin{align}
    \Delta \rho = \frac{\alpha_{\mathrm{QED}} \left( m_{Z}^{2} \right)}{16\pi^{2} m_{W}^{2} \left( 1 - m_{W}^{2}/m_{Z}^{2} \right)} &\left[ f\left( m_{H^{\pm}}^{2},m_{H}^{2} \right) + c_{\theta}^{2} \left( f\left( m_{H^{\pm}}^{2},m_{A}^{2} \right) - f\left( m_{A}^{2},m_{H}^{2} \right) \right)\right. \nonumber \\
    &\left. + s_{\theta}^{2} \left( f\left( m_{H^{\pm}}^{2},m_{a}^{2} \right) - f\left( m_{a}^{2},m_{H}^{2} \right) \right) \right]
\end{align}
with the function~$f$ being
\begin{align}
    f\left( x , y \right) = x+y-\frac{2xy}{x-y} \log \frac{x}{y} \, .
    \label{Eq:EWPTfunctionf}
\end{align}
Further constraints are those on the Peskin-Takeuchi parameters which are predicted by the SM to be
\begin{align}
    \mathcal{O}^{\mathrm{SM}} \equiv \left( S,T,U \right)^{\mathrm{SM}} = \left( 0.04,0.09,-0.02 \right)
\end{align}
and the experimentally allowed range is dictated by
\begin{align}
    \chi^{2} = \sum_{i,j} \left( \mathcal{O}_{i} - \mathcal{O}_{i}^{\mathrm{SM}} \right) \left( \sigma_{i}V_{ij}\sigma_{j} \right)^{-1} \left( \mathcal{O}_{j} - \mathcal{O}_{j}^{\mathrm{SM}} \right) 
\end{align}
with the standard deviation and symmetric covariance matrix
\begin{align}
    \sigma = \left( 0.11,0.14,0.11 \right) \quad , \quad V = \begin{pmatrix}
        1 & \times & \times \\
        0.92 & 1 & \times \\
        -0.68 & -0.87 & 1 
    \end{pmatrix} \, .
\end{align}
The mass differences between the fields from the singlet and second doublet affect the Peskin-Takeuchi parameters. In the alignment limit they can be written as~\cite{Haber:2010bw,Kanemura:2011sj,Arcadi:2023smv}
\begin{align}
    \mathcal{O}_{1} \equiv S &= -\frac{1}{4\pi} \left[ g \left( m_{H^{\pm}}^{2}, m_{H^{\pm}}^{2} \right) - c_{\theta}^{2}g \left( m_{H}^{2}, m_{A}^{2} \right) - s_{\theta}^{2}g \left( m_{H}^{2}, m_{a}^{2} \right) \right] \\
    \mathcal{O}_{2} \equiv T &= \frac{\Delta \rho}{\alpha_{\mathrm{QED}}} \\
    \mathcal{O}_{3} \equiv U &= -\frac{1}{4\pi} \left[ g \left( m_{H^{\pm}}^{2}, m_{H^{\pm}}^{2} \right) + c_{\theta}^{2}g \left( m_{H}^{2}, m_{A}^{2} \right) + s_{\theta}^{2}g \left( m_{H}^{2}, m_{a}^{2} \right) \right. \nonumber \\
    &\hspace{15mm} \left. - g \left( m_{H^{\pm}}^{2}, m_{H}^{2} \right) - c_{\theta}^{2}g \left( m_{H^{\pm}}^{2}, m_{A}^{2} \right) - s_{\theta}^{2}g \left( m_{H^{\pm}}^{2}, m_{a}^{2} \right) \right]
\end{align}
with the function
\begin{align}
    g \left( x, y \right) = -\frac{1}{3} \left( \frac{4}{3} - \frac{x \log x - y \log y}{x-y} - \frac{x+y}{\left( x-y \right)^{2}} f\left( x ,y \right) \right) \, .
    \label{Eq:EWPTfunctiong}
\end{align}

\clearpage

\section{Direct detection Wilson coefficients} \label{App:DirectdetectionWC}

This appendix aims to provide the technical details of the computation of the loop-induced DMDD cross section in the 2HDM+a and the matching between the parameters introduced in Section~\ref{Sec:2HDM_DMDD} and the Wilson coefficients of the \eDMEFT~(see Tab.~\ref{Tab:DDPmatch}).
\begin{table}[b!]
    \centering
    \begin{tabular}{llll}
    \toprule
        & \multicolumn{1}{c}{$\xi_\phi^q$} & \multicolumn{1}{c}{$\xi_{P_i}^\chi$} & \multicolumn{1}{c}{$g_{\phi P_i P_j}$} \\
        \midrule
        \multirow{3}*{$\{S_1,S_2\}=\{A,a\}$}  & \multirow{3}*{\shortstack{$\xi_h^q=(c_{q})^{(1)}_{0,0}$ \\ $\xi_H^q=0$}}& \multirow{3}*{\shortstack{$\xi_{A}^\chi=y_{1,0}^{(0)}$ \\ $\xi_{a}^\chi=y_{0,1}^{(0)}$}} & $g_{h A A}=\lambda_{2,0}^{(1)}\quad g_{H A A}=0$  \\
         & & &  $g_{h a A}=\lambda_{1,1}^{(1)}\quad g_{H a A}=0$\\
        & & &$g_{h a a}=\lambda_{0,2}^{(1)}\quad g_{H a a}=0$\\ \\
        \multirow{3}*{$\{S_1,S_2\}=\{H,a\}$}    & \multirow{3}*{\shortstack{$\xi_h^q=(c_{q})^{(1)}_{0,0}$ \\ $\xi_H^q=(c_{q})^{(0)}_{1,0}$}}& \multirow{3}*{\shortstack{$\xi_{A}^\chi=0$ \\ $\xi_{a}^\chi=y_{0,1}^{(0)}$}} & $g_{h A A}=0\quad g_{H A A}=0$  \\
         &  & &  $g_{h a A}=0\quad g_{H a A}=0$\\
        & & &$g_{h a a}=\lambda_{0,2}^{(1)}\quad g_{H a a}=\lambda_{1,2}^{(0)}$\\      
        \bottomrule
    \end{tabular}
    \caption{Relations between the Wilson coefficients and the couplings used for DMDD.}
    \label{Tab:DDPmatch}
\end{table}

Details about the original computation can be found in~\cite{Abe:2018emu}. As already pointed out in the aforementioned section, two main diagram topologies are to be taken into account, namely triangle and box diagrams. The Wilson coefficient associated to the former type and appearing in Eq.~(\ref{Eq:2HDM_DMDDWilsonCoefficients}) reads
\begin{align}
    C_{q}^{\mathrm{tri}} = -\sum_{\phi=h, H} \frac{\xi_{\phi}^{q}}{m_{\phi}^{2} v_h} C_{\phi \chi \chi}
\end{align}
with $\xi_{\phi}^{q}$ defined in Tab.~\ref{Tab:DDPmatch} for the matching scenarios considered and the last parameter given by
 \begin{align}
    C_{\phi \chi \chi} = - \frac{m_{\chi}}{16 \pi^{2}} &\left( g_{\phi a a} \left( \xi_{a}^{\chi} \right)^{2} \left[ \frac{\partial}{\partial p^{2}} B_{0} \left( p^{2}, m_{a}^{2}, m_{\chi}^{2} \right) \right]_{p^{2} = m_{\chi}^{2}} \right. \nonumber \\
    &\hspace{3mm} + \frac{2 g_{\phi a A} \xi_{A}^{\chi} \xi_{a}^{\chi}}{m_{A}^{2} - m_{a}^{2}} \mathcal{F}\left(m_\chi,m_A,m_a\right)  \nonumber \\
    &\hspace{3mm} \left. + g_{\phi A A} \left( \xi_{A}^{\chi} \right)^{2} \left[ \frac{\partial}{\partial p^{2}} B_{0} \left(p^{2}, m_{A}^{2}, m_{\chi}^{2} \right) \right]_{p^{2} = m_{\chi}^{2}} \right) \, ,
 \end{align}
where $g_{\phi i j}$ are the couplings between the scalar mediators~$\phi \in \{ h,H \}$ and the pseudoscalars $P_i \in \{A,a\}$ and $\xi_{P_i}^{\chi}$ are the couplings between the DM particle and the pseudoscalar~$P_i$. Following the result from~\cite{Abe:2018emu}, the function~$B_{0}$ is defined as
\begin{align}
\left. \left[ \frac{\partial}{\partial p^{2}}B_{0} \left( p^{2}, m_{1}, m_{\chi}^{2} \right) \right] \right \vert_{p^{2} = m_{\chi}^{2}} &= \left. \int_{0}^{1} \dd x \ \frac{x\left( 1-x \right)}{m_{1}^{2} x + m_{\chi}^{2} \left( 1-x \right) - p^{2} x \left( 1-x \right)} \right\vert_{p^{2} = m_{\chi}^{2}} \nonumber \\
&= - \frac{1}{m_{\chi}^{2}} + \frac{\left( m_{1}^{2} - m_{\chi}^{2} \right) \log \left( \frac{m_{1}^{2}}{m_{\chi}^{2}} \right)}{2 m_{\chi}^{4}} \nonumber \\
&+ \frac{2 \sqrt{m_{1}^{4} - 4 m_{1}^{2} m_{\chi}^{2}} \left(m_{1}^{2} - 3 m_{\chi}^{2} \right) \log \left( \frac{\sqrt{m_{1}^{4} - 4 m_{1}^{2} m_{\chi}^{2}} + m_{1}^{2}}{2 m_{1} m_{\chi}} \right)}{8 m_{\chi}^{6} - 2 m_{1}^{2} m_{\chi}^{4}}
\end{align}
and the function $\mathcal{F}$ is given by
\begin{align}
&\mathcal{F}\left(m_\chi,m_1,m_2\right) = B_{1} \left( m_{\chi}^{2}, m_{1}^{2}, m_{\chi}^{2} \right) - B_{1} \left( m_{\chi}^{2}, m_{2}^{2}, m_{\chi}^{2} \right)\nonumber\\
&= \int_{0}^{1} \dd x \ \left( 1-x \right)  \log \frac{m_{1}^{2} x + m_{\chi}^{2} \left( 1-x \right) - m_\chi^{2} x \left( 1-x \right)}{m_{2}^{2} x + m_{\chi}^{2} \left( 1-x \right) - m_\chi^{2} x \left( 1-x \right)}  \nonumber \\
&= \frac{m_1^2 \left[\left(m_1^2-2 m_{\chi }^2\right) \log
    \left(\frac{m_1^2}{m_{\chi }^2}\right)-2 \left\{ m_{\chi }^2+m_1
    \sqrt{m_1^2-4 m_{\chi }^2} \log \left(\frac{\sqrt{m_1^2-4 m_{\chi
    }^2}+m_1}{2 m_{\chi }}\right)\right\}\right]}{4 m_{\chi }^4}\nonumber \\
&- \frac{m_2^2 \left[\left(m_2^2-2 m_{\chi }^2\right) \log
    \left(\frac{m_2^2}{m_{\chi }^2}\right)-2 \left\{ m_{\chi }^2+m_2
    \sqrt{m_2^2-4 m_{\chi }^2} \log \left(\frac{\sqrt{m_2^2-4 m_{\chi
    }^2}+m_2}{2 m_{\chi }}\right)\right\}\right]}{4 m_{\chi }^4}
 \end{align}
The gluon contribution reads
\begin{align}
    C_G^{\mathrm{tri}} = \frac{2}{27} \sum_{Q=c, b, t} C_{Q}^{\mathrm{tri}} \, .
\end{align}

 For the case of the box diagrams the coefficients are
 \begin{align}
 \nonumber
 C_q^{\text {box }}=\frac{-m_\chi}{(4 \pi)^2}\left(\frac{m_q}{v}\right)^2\bigg\{ & \frac{\left(\xi_a^\chi \xi_a^q\right)^2}{m_a^2}\left[G\left(m_\chi^2, 0, m_a^2\right)-G\left(m_\chi^2, m_a^2, 0\right)\right] \\
 \nonumber
 & +\frac{\left(\xi_A^\chi \xi_A^q\right)^2}{m_A^2}\left[G\left(m_\chi^2, 0, m_A^2\right)-G\left(m_\chi^2, m_A^2, 0\right)\right] \\
 & +2 \frac{\xi_A^\chi \xi_a^\chi \xi_A^q \xi_a^q}{m_A^2-m_a^2}\left[G\left(m_\chi^2, m_A^2, 0\right)-G\left(m_\chi^2, m_a^2, 0\right)\right]\bigg\}
 \end{align}
 \begin{align}
 \nonumber
 C_q^{(1) \mathrm{box}}=\frac{-8}{(4 \pi)^2}\left(\frac{m_q}{v}\right)^2\bigg\{ & \frac{\left(\xi_a^\chi \xi_a^q\right)^2}{m_a^2}\left[X_{001}\left(m_\chi^2, m_\chi^2, 0, m_a^2\right)-X_{001}\left(m_\chi^2, m_\chi^2, m_a^2, 0\right)\right] \\
 \nonumber
 &\hspace{-30mm} +\frac{\left(\xi_A^\chi \xi_A^q\right)^2}{m_A^2}\left[X_{001}\left(m_\chi^2, m_\chi^2, 0, m_A^2\right)-X_{001}\left(m_\chi^2, m_\chi^2, m_A^2, 0\right)\right] \\
 &\hspace{-30mm} +2 \frac{\xi_A^\chi \xi_a^\chi \xi_A^q \xi_a^q}{m_A^2-m_a^2}\left[X_{001}\left(m_\chi^2, m_\chi^2, m_A^2, 0\right)-X_{001}\left(m_\chi^2, m_\chi^2, m_a^2, 0\right)\right] \bigg\}
 \end{align}
 \begin{align}
 \nonumber
 C_q^{(2) \mathrm{box}}=\frac{-4 m_\chi}{(4 \pi)^2}\left(\frac{m_q}{v}\right)^2\bigg\{& \frac{\left(\xi_a^\chi \xi_a^q\right)^2}{m_a^2}\left[X_{111}\left(m_\chi^2, m_\chi^2, 0, m_a^2\right)-X_{111}\left(m_\chi^2, m_\chi^2, m_a^2, 0\right)\right] \\
 \nonumber
 &\hspace{-30mm} +\frac{\left(\xi_A^\chi \xi_A^q\right)^2}{m_A^2}\left[X_{111}\left(m_\chi^2, m_\chi^2, 0, m_A^2\right)-X_{111}\left(m_\chi^2, m_\chi^2, m_A^2, 0\right)\right] \\
 &\hspace{-30mm} +2 \frac{\xi_A^\chi \xi_a^\chi \xi_A^q \xi_a^q}{m_A^2-m_a^2}\left[X_{111}\left(m_\chi^2, m_\chi^2, m_A^2, 0\right)-X_{111}\left(m_\chi^2, m_\chi^2, m_a^2, 0\right)\right] \bigg\} \, ,
 \end{align}
where
 \begin{equation}
 G\left(m_\chi^2, m_1^2, m_2^2\right)=6 X_{001}\left(m_\chi^2, m_\chi^2, m_1^2, m_2^2\right)+m_\chi^2 X_{111}\left(m_\chi^2, m_\chi^2, m_1^2, m_2^2\right)
 \end{equation}
 \begin{align}
X_{001}\left(p^2, M^2, m_1^2, m_2^2\right)=\int_0^1 d x \int_0^{1-x} d y \frac{\frac{1}{2} x(1-x-y)}{M^2 x+m_1^2 y+m_2^2(1-x-y)-p^2 x(1-x)}
 \end{align}
 \begin{align}
X_{111}\left(p^2, M^2, m_1^2, m_2^2\right)=\int_0^1 d x \int_0^{1-x} d y \frac{-x^3(1-x-y)}{\left[M^2 x+m_1^2 y+m_2^2(1-x-y)-p^2 x(1-x)\right]^2}
 \end{align}
The WC for the gluon's box diagrams is given by
 \begin{align}
 C_G^{\mathrm{box}}=\sum_{Q=c, b, t} \frac{-m_\chi}{432 \pi^2}\left(\frac{m_Q}{v}\right)^2\left[\left(\xi_a^\chi \xi_a^Q\right)^2 \frac{\partial F\left(m_a^2\right)}{\partial m_a^2}\right. & +\left(\xi_A^\chi \xi_A^Q\right)^2 \frac{\partial F\left(m_A^2\right)}{\partial m_A^2} \\
 & \left.+2 \xi_A^\chi \xi_a^\chi \xi_A^Q \xi_a^Q \frac{\left[F\left(m_A^2\right)-F\left(m_a^2\right)\right]}{m_A^2-m_a^2}\right] \, ,
 \end{align}
 where
 \begin{align}
 F\left(m_a^2\right)=\int_0^1 d x\{ & 3 Y_1\left(p^2, m_\chi^2, m_a^2, m_Q^2\right) \\
 & -m_Q^2 \frac{\left(2+5 x-5 x^2\right)}{x^2(1-x)^2} Y_2\left(p^2, m_\chi^2, m_a^2, m_Q^2\right) \\
 & \left.-2 m_Q^4 \frac{\left(1-2 x+2 x^2\right)}{x^3(1-x)^3} Y_3\left(p^2, m_\chi^2, m_a^2, m_Q^2\right)\right\} \, ,
 \end{align}
where the definitions of the loop functions $Y_1, Y_2$, and $Y_3$ and  $\partial F\left(m_a^2\right) / \partial m_a^2$ are obtained from~\cite{Abe:2018emu}.

\clearpage

\end{appendix}
  
\clearpage

\bibliographystyle{JHEP}    
\bibliography{nLEFT}

\end{document}